\documentclass[twocolumn]{aastex6}
\usepackage{graphicx}
\usepackage{array}
\usepackage{amssymb}
\usepackage{natbib}
\usepackage{float}
\usepackage{color}
\usepackage[utf8]{inputenc}  %Additional package for equation
\usepackage{amsmath}  %addition package for math
\usepackage{comment}
\usepackage{footnote}
\usepackage{longtable}
\usepackage{url}
\usepackage[caption2]{ccaption}
\usepackage[figuresright]{rotating}
\usepackage{chngcntr}
\newcommand{\degree}{\ensuremath{^\circ}}
\renewcommand*{\thefootnote}{\fnsymbol{footnote}}
\begin{document}


\title{A catalog of Narrow line Seyfert 1 galaxies from the Sloan Digital Sky Survey data release 12}

%% Use \author, \affil, plus the \and command to format author and affiliation 
%% information.  If done correctly the peer review system will be able to
%% automatically put the author and affiliation information from the manuscript
%% and save the corresponding author the trouble of entering it by hand.
%%
%% The \affil should be used to document primary affiliations and the
%% \altaffil should be used for secondary affiliations, titles, or email.

%% Authors with the same affiliation can be grouped in a single
\author{Suvendu Rakshit \altaffilmark{1,2}, C.S. Stalin \altaffilmark{1}, Hum Chand \altaffilmark{3} and Xue-Guang Zhang \altaffilmark{4}}


\altaffiltext{1}{Indian Institute of Astrophysics, Block II, Koramangala, Bangalore-560034, India}
\altaffiltext{2}{suvenduat@gmail.com}
\altaffiltext{3}{Aryabhatta Research Institute of Observational Sciences (ARIES), 263002, Nainital, India}
\altaffiltext{4}{Institute of Astronomy and Space Science, Sun Yat-Sen University, No. 135, Xingang Xi Road, Guangzhou, 510275, P. R. China}
\shorttitle{NLSy1 galaxies from SDSS DR12}
\shortauthors{Rakshit et al.}
\slugcomment{2017, ApJS, 229, 39R}
\begin{abstract}
We present a new catalog of narrow-line Seyfert 1 (NLSy1) galaxies from the Sloan Digital Sky Survey Data Release 12 (SDSS DR12). This was obtained by a systematic analysis through modeling of the continuum and emission lines of the spectra of all the 68,859 SDSS DR12 objects that are classified as ``QSO'' by the SDSS spectroscopic pipeline with $z<0.8$ and a median signal-to-noise ratio (S/N) $>2$ pixel$^{-1}$. This catalog contains a total of 11,101 objects, which is about five times larger than the previously known NLSy1 galaxies. Their monochromatic continuum luminosity at 5100 $\mathrm{\AA}$  is found to be strongly correlated with H$\beta$, H$\alpha$ and [O {\small III}] emission line luminosities. The optical Fe {\small II} strength in NLSy1 galaxies is about two times larger than the broad-line Seyfert 1 (BLSy1) galaxies.  About 5\% of the catalog sources are detected in FIRST survey. The Eddington ratio ($\xi_{\mathrm{Edd}}$) of NLSy1 galaxies has an average of $\log \, \xi_{\mathrm{Edd}}$ of $-0.34$, much higher than $-1.03$ found for BLSy1 galaxies. Their black hole masses ($M_{\mathrm{BH}}$) have an average $\log \, M_{\mathrm{BH}}$ of $6.9 \, M_{\odot}$, which is less than BLSy1 galaxies, which have an average of $\log \, M_{\mathrm{BH}}$ of $8.0 \, M_{\odot}$. The $M_{\mathrm{BH}}$ of NLSy1 galaxies is found to be correlated with their host galaxy velocity dispersion. Our analysis suggests that geometrical effects playing an important role in defining NLSy1 galaxies and their $M_{\mathrm{BH}}$ deficit is perhaps due to their lower inclination compared to BLSy1 galaxies.
\end{abstract}




%======================================================
%======================  Section ======================
%======================================================

\section{Introduction}
Seyfert galaxies, a type of active galactic nuclei (AGNs) are generally classified into 
two broad categories depending on their emission line properties; namely Seyfert 1 (Sy 1)
and Seyfert 2 (Sy 2) galaxies \citep[see][]{2015ARA&A..53..365N}. Traditionally, an AGN is termed a Seyfert galaxy or a quasar when its absolute optical magnitude is fainter or brighter than a (somewhat arbitrary) value, e.g., $M_B = -23$ \citep{2010A&A...518A..10V}.  
Sy 1 galaxies show both broad permitted emission lines that originate from the broad-line region (BLR) having widths of a few thousand 
km s$^{-1}$ and narrow forbidden emission lines that originate from the narrow line 
region (NLR) and  having widths of a few hundred km s$^{-1}$. On the other hand, Sy 2 galaxies show narrow 
permitted and forbidden lines in their emission line spectra \citep{1996agn..book.....R}. The varied observational differences seen between 
Sy 1 and Sy 2 galaxies are explained by unification model
\citep{1993ARA&A..31..473A}, according to which, both of them have similar 
internal structure and the differences in their spectra are mainly due to 
orientation effects. %According to this simple unification model,
The narrow lines seen in the spectra of Sy 2 galaxies are due to the fact that they are viewed  
close to edge-on and because the view of their BLR is obscured by the torus. However, based on recent observations, \citet{2015ARA&A..53..365N} concluded that the simple unification scheme requires some major modifications.  

Although the division between Seyfert galaxies is well defined, exceptions 
have been found. Some Seyfert galaxies show narrow ``broad 
permitted lines'' similar to the Sy 2 galaxies, though they have all the spectral properties of Sy 1 sources. They were classified as 
narrow-line Seyfert 1 (NLSy1) galaxies by \citet{1985ApJ...297..166O}, initially 
based on the presence of narrow permitted emission lines and weak 
[O {\small III}] lines relative to H$\beta$ with 
[O {\small III}]$\lambda 5007/ \mathrm{H}\beta<3$. Later, \citet{1989ApJ...342..224G} put an upper limit on the width of the permitted 
lines, the full width at half maximum (FWHM) of the broad H$\beta$ line $<2000\, \mathrm{km\,s^{-1}}$, to quantitatively define 
the NLSy1 category. NLSy1 galaxies often show strong Fe {\small II} emission lines 
relative to H$\beta$ in the UV and optical spectral domain, a strong 
soft X-ray excess, and high amplitude rapid X-ray variability 
\citep{1996A&A...305...53B,1999ApJS..125..317L}.
In the soft X-ray band, NLSy1 galaxies have a photon index of 
$\Gamma$= $2.19\pm 0.10$ \citep{1999ApJS..125..317L}, which is significantly steeper than the broad-line Seyfert 1 (BLSy1)
galaxies \citep{1994MNRAS.268..405N}. They are  generally believed to have
lower black hole masses (10$^6$ - 10$^8 \,M_{\odot}$) and higher Eddington 
ratios than BLSy1 galaxies \citep[hereafter ZH06]{2006ApJS..166..128Z}
since the former have narrower Balmer lines than the 
latter \citep[see also][]{2012AJ....143...83X}.  %\citep{2006ApJS..166..128Z,2012AJ....143...83X}. 
However, there are reports that the black hole masses of NLSy1 and 
BLSy1 galaxies
are indifferent, and geometrical effects can fully account for the mass deficit in
NLSy1 galaxies \citep{2013MNRAS.431..210C,2016MNRAS.458L..69B,2016IJAA....6..166L}. A majority of the population of NLSy1 galaxies
are radio quiet with a minority of about 7\% are known to be radio-loud \citep{2000NewAR..44..381P}. The radio loud fraction of NLSy1 galaxies 
is indeed half of the 15\% that we know for the quasar category of AGNs \citep{1989AJ.....98.1195K}.

One of the many correlations that drive the eigenvector 1 (E1) in AGNs is the strong anti-correlation between Fe {\small II} emission and [O {\small III}] line 
strength \citep{1992ApJS...80..109B,1999ApJ...514...40M,1999A&A...350..805G}. 
Interestingly, NLSy1 galaxies lie at the extremely low [O {\small III}] or negative end 
of E1, providing a unique opportunity to study the physical parameters 
responsible for such an unusual class. The drivers of E1 are expected to be 
the black hole mass, accretion rate, orientation angle, and covering factor or 
anisotropy of the BLR \citep{1992ApJS...80..109B,1996A&A...309...81W,2002ApJ...565...78B}. Recently, 
\citet{2014Natur.513..210S} showed that the main driving physical parameters 
of E1 in AGNs are the Eddington ratio and orientation that affect the observed line profile.

The number of NLSy1 galaxies has gradually increased in the last two decades. From the Solan Digital Sky Survey (SDSS) early data release, \citet{2002AJ....124.3042W} compiled a catalog of 150 NLSy1 galaxies. This number increased to 2011 sources, when
SDSS DR3 was used by ZH06. There is an about a factor of 10 increase in the number of AGNs in SDSS spectroscopic 
data release 12 \citep[SDSS DR12][]{2015ApJS..219...12A} than in DR3 offering the identification of a large sample of new NLSy1 galaxies. The main motivation for this
work is therefore to increase the number of NLSy1 galaxies we know today, using the enlarged list of sources from SDSS DR12. The finding
of a large sample of NLSy1 galaxies from a systematic analysis of SDSS DR12
spectra could allow us to study their properties, in general, and to verify the previously known correlations and establish new correlations on 
various physical properties of these sources. Toward this goal, we have 
systematically searched for NLSy1 galaxies in the spectroscopic ``QSO'' 
sample of SDSS DR12. In this work,  
we present this new catalog of NLSy1 galaxies and the results of various  
analyses carried out on this new sample to understand their properties.  
This paper is organized as follows. In section \ref{sec:analysis}, we present the data sets used for searching NLSy1 galaxies and describe the spectral analysis procedure. The sample of NLSy1 galaxies is presented in 
section \ref{sec:sample} and their properties are discussed in 
section \ref{sec:discussion}. A summary of the key results of the work is given in 
section \ref{sec:summary}. A cosmology with $H_0 = 70\, \mathrm{km\,s^{-1}\, Mpc^{-1}}$, $\Omega_\mathrm{m} = 0.3$, and $\Omega_{\lambda} = 0.7$ is assumed throughout.

\begin{figure*}
\centering 
\resizebox{18cm}{18.0cm}{\includegraphics{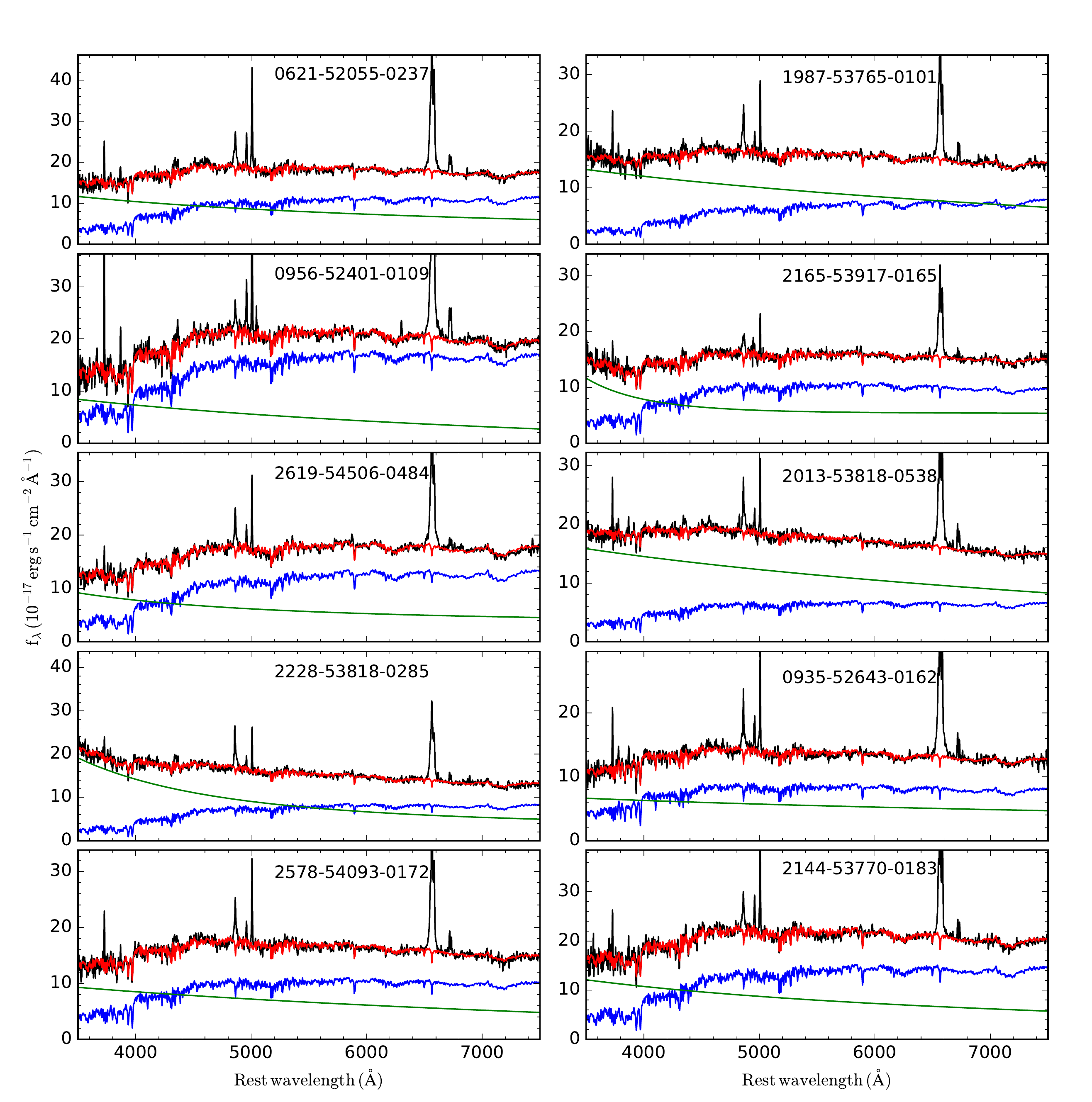}}
\caption{Subtraction of stellar contribution from spectra. Here, the black line shows the  observed spectrum, the red line shows the best-fitted results, the green line shows power-law continuum fit and the blue line shows the stellar contribution. The labels indicate corresponding PLATE-MJD-FIBER of SDSS. The observed spectra were smoothed by a boxcar of $5$ pixels only for illustration.}\label{Fig:con_fit}.
\end{figure*}

%======================================================
%======================  Section ======================
%======================================================

\section{Data and spectral analysis}\label{sec:analysis}
We selected objects from the ``specObj''\footnote{``specObj'' contains only the best spectra for any object obtained by SDSS (see \url{http://www.sdss.org/dr12/help/faq/\#scienceprimary}).} data product of the SDSS DR12 catalog \citep{2015ApJS..219...12A} that are classified as ``QSO'' by the SDSS spectroscopic pipeline \citep{2002AJ....123.2945R,2012ApJS..199....3R}. This can include both SDSS and Baryon Oscillation Spectroscopic Survey \citep[BOSS;][]{2013AJ....145...10D} spectra with a wavelength coverage of $\lambda3800-9200 \mathrm{\AA}$ and $\lambda3600-10400\mathrm{\AA}$ respectively. However, to ensure the presence of both H$\beta$ and [O {\small III}] emission lines within the wavelength coverage of SDSS and BOSS spectra as well as for the convenience of comparison to ZH06, we considered objects with $z<0.8$. This yielded 114,806 sources, of which only sources that have a median signal-to-noise ratio (S/N) $>$ 2 pixel$^{-1}$ were retained. This resulted in 68,859 sources that were taken further for a detailed spectral 
analysis. 

%\section{Spectral analysis}\label{sec:analysis}
To carefully analyze the spectra and accurately estimate the emission line parameters, we performed a two-step fitting process \citep[see also,][]{2014MNRAS.438..557Z,2016MNRAS.457.3878Z} after shifting the spectra to their rest frame.
\begin{enumerate}
\item[I.] The spectra were simultaneously fitted with an AGN power-law continuum (global) and stellar contribution of host galaxy using Eq. \ref{eq:SSP}. During this step, we masked the AGN emission lines except Fe {\small II} multiplets (see section \ref{sec:host}). We then subtracted the stellar contribution from the spectra without subtracting the AGN global continuum, and proceeded to the next step given below.     
\item[II.] In this step, we simultaneously fitted AGN emission lines including local AGN continuum along with Fe {\small II} template to the host galaxy subtracted spectra obtained in Step I above (see section \ref{sec:fitting}). From this fitting, we estimated all the emission line and continuum parameters of AGNs.     
\end{enumerate}

%======================  Subsection ======================

\subsection{Subtraction of host contribution}\label{sec:host}
The spectra of low-redshift AGNs may contain a significant 
contribution from the host galaxy starlight. To elucidate important
information on AGN properties from the spectra, the stellar light contribution
needs to be removed. We used the simple stellar population (SSP) method
to remove the contribution of host galaxy from the spectra and
estimated the stellar properties following \citet{2014MNRAS.438..557Z} and 
\citet{2016MNRAS.457.3878Z}. The observed spectrum, $F(\lambda)$, can be 
modeled using SSP method as follows.
\begin{equation}
F(\lambda) =  \left[ \sum_{i=1}^{39} a_i \times F_{\mathrm{ssp}} (\delta \lambda, r_{\lambda})  \right]  \ast g(\lambda, \sigma_{*})  + F_{\mathrm{AGN}}(\lambda, r_{\lambda})
\label{eq:SSP}
\end{equation} 
where the first part of the right-hand side represents the contribution of host galaxy starlight and the second part represents the AGN contribution. Here $a_i$ is the amplitude of the individual SSP templates, $ F_{\mathrm{ssp}} (\delta \lambda, r_{\lambda})$ is the SSP template with a wavelength shift of $\delta \lambda$ and reddening factor of $r_{\lambda}$, $g(\lambda, \sigma_{*})$ is the Gaussian broadening function with velocity $\sigma_{*}$ considered as the stellar velocity dispersion, and $F_{\mathrm{AGN}}(\lambda, r_{\lambda})$ is the contribution from the AGN in the form of $\lambda^{\alpha}$ with the reddening correction factor $r_{\lambda}$ ($\alpha$ is the power-law index). The summation runs from $i=1$ to 39 as the SSP model is applied with 39 simple stellar template spectra. The templates were taken from \citet{2003MNRAS.344.1000B} having ages between 5 Myr and 12Gyr  and solar metallicities of $Z=0.008$, 0.05, and 0.02.  A detailed description of the SSP method is given in \citet{2003MNRAS.344.1000B}, \citet{2003MNRAS.346.1055K}, \citet{2009ApJ...705L..76W}, \citet{2012Natur.484..485C}, and \citet{2014MNRAS.438..557Z}.

We masked the emission lines without considering the Fe {\small II} multiplets as emission lines so as to leave  sufficient continuum windows for continuum fitting and performed Levenberg$-$Marquardt 
least-squares minimization using \textit{IDL mpfit}\footnote{\url{http://purl.com/net/mpfit}} fitting package 
\citep{2009ASPC..411..251M} to fit the spectra. This allowed us to decompose the 
contribution of the host galaxy and AGN continuum from the spectra. We then subtracted the contribution of the host galaxy stellar light from the spectra without subtracting AGN global continuum. Figure 
\ref{Fig:con_fit} shows examples of starlight subtraction from the spectra. The decomposed AGN continuum emission component is shown by a solid green line, host galaxy contribution is shown by a solid blue line. We note that the continuum fitting procedure is phenomenological, which is sufficient for the present study.

\begin{figure*}
\centering
\resizebox{14cm}{20.0cm}{\includegraphics{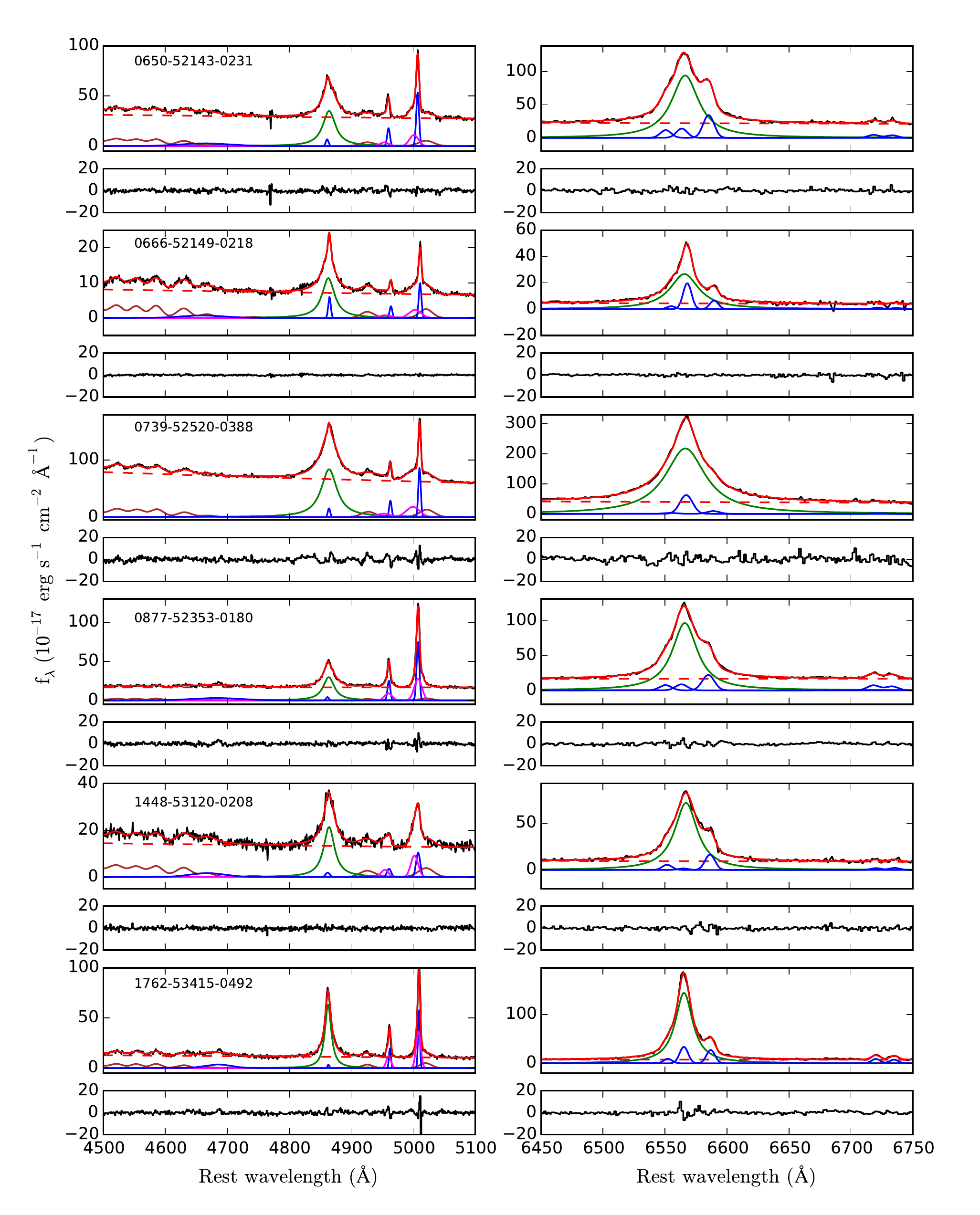}}
\caption{Examples of our simultaneous spectral fitting around H$\beta$ (left) and H$\alpha$ (right) emission line regions. The labels are PLATE-MJD-FIBER of SDSS. In each panel, the best-fitting results are shown in the top and the corresponding residual plots (horizontal black line) are shown below. In the H$\beta$ region (left), the observed spectrum (black) and the overall fitted spectrum (red) are shown with the decomposed individual components; broad H$\beta$ in green, narrow components (H$\beta_n$, [O {\small III}] doublet) and broad He {\small II} line in blue, broad [O {\small III}] doublet in magenta, and Fe {\small II} lines in brown. In the H$\alpha$ region (right), the observed spectrum (black) and the overall fitted spectrum (red) is shown with the decomposed individual components; broad H$\alpha$ in green and narrow components (H$\alpha_n$, [N {\small II}], [S {\small II}] doublets) in blue. Note that the x-scales of the figures are limited for the purpose of visualization.}\label{Fig:spectra}. 
\end{figure*}

%======================  Subsection ======================

\subsection{Emission line fitting}\label{sec:fitting}
After subtracting the starlight contribution from the spectra as described above, we performed a simultaneous fitting of emission lines including one local power-law continuum component around H$\beta$ and H$\alpha$ regions along with recent high-quality Fe {\small II} template spectra taken from \citet{2010ApJS..189...15K}. In the H$\beta$ region, with a wavelength range of $4385\mathrm{\AA} - 5500\, \mathrm{\AA}$, 
apart from H$\beta$ (having both narrow and broad components) the lines fitted are He {\small II}$\lambda 4687 \,\mathrm{\AA}$ and [O {\small III}]$\lambda 4959, \,5007 \, \mathrm{\AA}$ doublet (with two Gaussian 
functions). In the H$\alpha$ region, with a wavelength range $6280\,\mathrm{\AA} - 6750\mathrm{\AA}$,  the lines used in the 
fitting process are the narrow [O {\small I}]$\lambda 6300,\,6363 \,\mathrm{\AA}$, 
narrow [N {\small II}]$\lambda 6548, \,6583 \,\mathrm{\AA}$ doublet, and the 
narrow [S {\small II}]$\lambda 6716, \,6731 \,\mathrm{\AA}$ doublet along with H$\alpha$. Each of the narrow components are fitted with a single Gaussian having a maximum FWHM of 1200 km/s following \citet{2011ApJS..194...45S}. From theoretical arguments, \citet{2012MNRAS.426.3086G} suggest that 
the Balmer lines of NLSy1 galaxies are expected to have Lorentzian profile caused by the 
microscopic turbulence velocity of BLR clouds enhancing the line wings 
relative to the core, especially at a large radius. Observationally,
it has also been noticed that the observed broad Balmer lines are best characterized 
by a Lorentzian than a Gaussian profile \citep{1996ApJS..106..341M,2001A&A...372..730V,2002ApJ...566L..71S,2016arXiv160703438C}.

In our fitting process, when both H$\beta$ and H$\alpha$ regions are fitted simultaneously, we allow a single Gaussian and Lorentzian functions for fitting both the broad Balmer components, where the algorithm automatically chooses among these functions the best-fitted function, which gives a similar width for both broad components of Balmer lines, H$\beta$ and H$\alpha$. We also find that, in our fitting, the Lorentzian function is preferred compared to the Gaussian function, confirming the earlier claims by various authors as discussed above. This led us to choose a single Lorentzian function to model the broad component of H$\beta$ when only the H$\beta$ region was fitted. Note that if the narrow H$\beta$/H$\alpha$ component is not required with the broad component, our iterative fitting procedure automatically drops it during minimization.  

Spectral coverage of SDSS and BOSS allows us to carry out a simultaneous fitting of both H$\alpha$ and H$\beta$ regions for $z<0.3629$, and for higher $z$ (up to 0.8) only fitting of H$\beta$ region was performed. During the fitting, the flux ratios of 
[O {\small III}] and [N {\small II}] doublets were fixed to their theoretical 
values, i.e. $F(5007)/F(4959) = 3$ and $F(6585)/F(6549) = 3$. Widths of all the 
narrow components in the H$\beta$ region were tied with the narrow [O {\small III}] line width and the redshift of each doublet was tied together \citep[see][]{2011ApJS..194...45S}. Similarly, in the H$\alpha$ region, widths of all the narrow components were tied together 
with narrow [N {\small II}]$\lambda \,6583\mathrm{\AA}$. A final fitting was carried out by simultaneously varying all the free parameters using Levenberg$-$Marquardt least-squares minimization, 
allowing us to estimate all emission line parameters including Fe {\small II} 
parameters along with the nuclear continuum contribution. Figure \ref{Fig:spectra} 
shows few examples of emission line fitting around H$\beta$ (left) and 
H$\alpha$ (right). The corresponding residual plots are shown in the lower panels.

\section{Criteria to select NLSy1 galaxies}\label{sec:sample}
Using the spectral fitting procedure described above, line parameters for all the 68,859 sources were estimated that were then used to find genuine NLSy1 galaxies provided they fulfill all of the following four criteria \citep{1985ApJ...297..166O,1989ApJ...342..224G,2006ApJS..166..128Z}.

\begin{enumerate}
\item The line flux of the broad component of H$\beta$\footnote{Hereafter, the line flux and width 
is of the broad component, unless stated otherwise.} line is detected at more than a $3\sigma$ level.
 
\item The width of the broad H$\beta$ component is greater than the width of the narrow forbidden line.  
\item The FWHM of the broad H$\beta$ emission line is narrower than 2200 $\mathrm{km\, s^{-1}}$.  
\item The flux ratio of total [O {\small III}] to 
total H$\beta <3$, where total refers to the sum of both broad and narrow components flux.
\end{enumerate}
Adopting the above criteria lead to the identification of 11,222 NLSy1 
galaxies. All the fitting results were visually inspected and 217 appears to 
have large scatter in FWHM(H$\beta$)/FWHM(H$\alpha$) than others, mainly due to low S/N
spectra. Those 217 spectra were fitted again using an additional constraint, 
FWHM(H$\alpha$)=0.9 $\times$ FWHM(H$\beta$), which is the relation obtained by \citet{2016MNRAS.457.3878Z}. After the second fit, out of these 217 
objects, 96 satisfy the condition of NLSy1 galaxies making a final sample of 11,101 NLSy1 galaxies. The details of the final list of sources
are given in Table \ref{Table:object}.

%======================================================
%======================  Section ======================
%======================================================

We note that the line width cutoff considered here is H$\beta \le 2200\, 
\mathrm{km\, s^{-1}}$, instead of the original value of 
H$\beta<2000 \, \mathrm{km\, s^{-1}}$ put by \citet{1989ApJ...342..224G}. This criterion was taken following ZH06 and considering the fact that the line width 
distribution of broad Balmer lines do not show any sharp cutoff between BLSy1 
and NLSy1 population and rather shows a smooth distribution of width. Out of 
the 11,101 NLSy1 galaxies, 8577 have FWHM(H$\beta$)$<2000 \, \mathrm{km\, s^{-1}}$. 
%{Interestingly, broad Balmer components of about 2999  (75 \%) objects out of 
%4021 NLSy1 for which both Gaussian and Lorentzian profiles were allowed to 
%fit broad components i.e., objects having $z\le 0.3629$, were best fitted by 
%a Lorentzian while only 1022 (25 \%) objects were best fitted by Gaussian 
%profile for objects. 
For a majority of the sources in our sample, the Balmer lines were best-fitted with a Lorentzian than a Gaussian profile.
This reaffirms the claim made in the literature, as discussed in section \ref{sec:fitting}, that the broad Balmer lines of NLSy1 galaxy can be best-fitted by 
a single Lorentzian.

\begin{figure}
\centering
\resizebox{8cm}{7.0cm}{\includegraphics{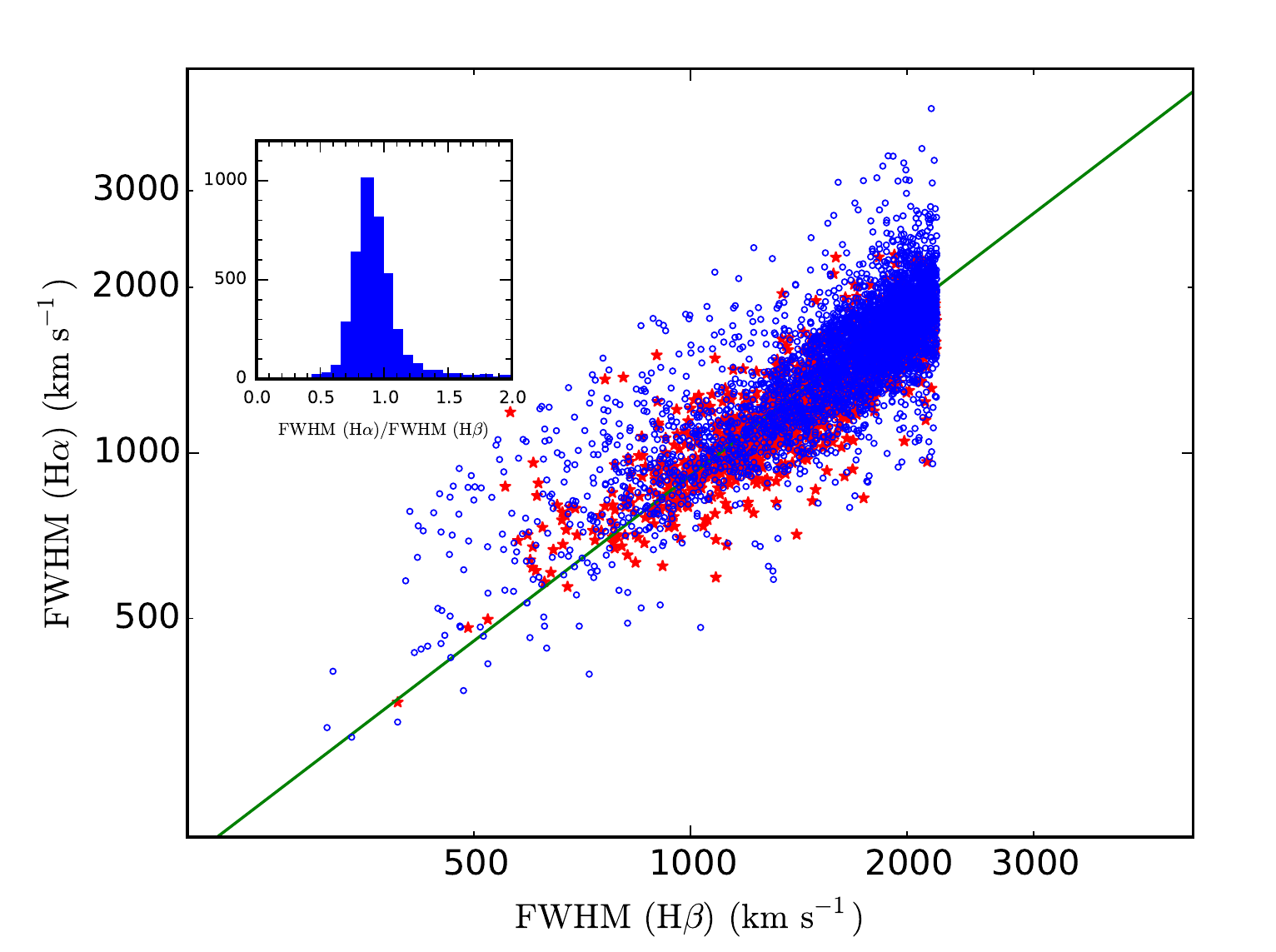}}
\caption{Correlation between FWHM of H$\alpha$ and H$\beta$ of 4021 NLSy1 galaxies in which both H$\alpha$ and H$\beta$ lines have been detected in our sample (circle) and the 1220 NLSy1 of ZH06 sample (star). The solid green line shows the best linear fit $\mathrm{FWHM}(\mathrm{H}\alpha)=(0.909 \pm 0.002) \times \mathrm{FWHM}(\mathrm{H}\beta)$. The inset shows the histogram of the FWHM ratio of H$\alpha$ to H$\beta$.}\label{Fig:fwhm_hahb}. 
\end{figure}

\begin{figure}
\centering
\resizebox{8cm}{7.0cm}{\includegraphics{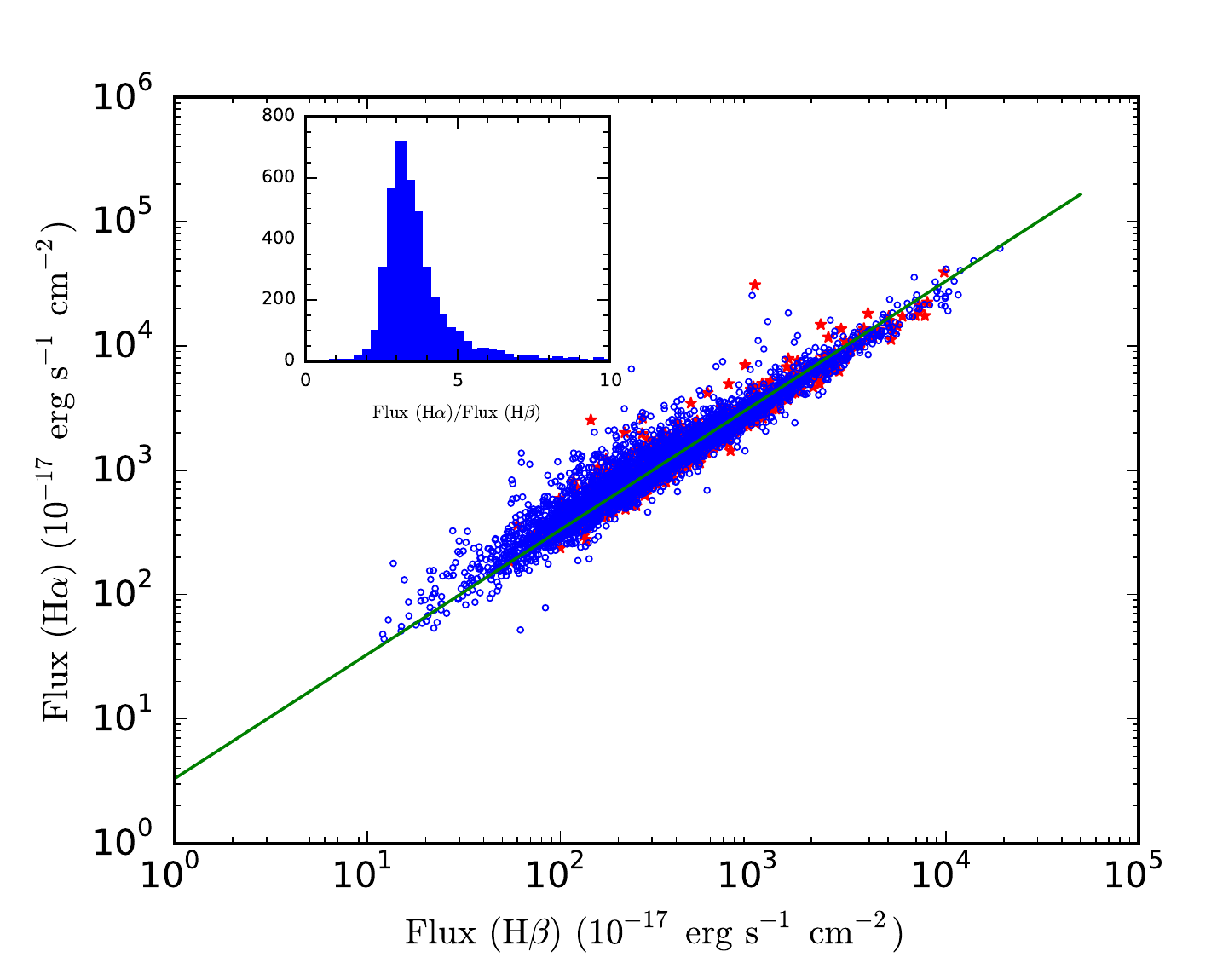}}
\caption{Correlation between the flux of H$\alpha$ and H$\beta$ of all the NLSy1 galaxies shown in Figure \ref{Fig:fwhm_hahb}. The solid line shows the best linear fit $F(\mathrm{H}\alpha) = (3.312 \pm 0.012) \times \mathrm{F(H\beta)}$. The inset shows the histogram of the flux ratio of H$\alpha$ to H$\beta$. }\label{Fig:flux_hahb}. 
\end{figure}

We have plotted in Figure \ref{Fig:fwhm_hahb}, the FWHM of broad Balmer 
components for 4021 NLSy1 galaxies (dots) having $z$ $<$ 0.3629, so
that both the Balmer lines are in the SDSS wavelength coverage. The stars 
in the figure indicate all of the 1220 NLSy1 galaxies from the ZH06 sample for comparison. 
We found $\mathrm{FWHM}(\mathrm{H}\alpha)=(0.909 \pm 0.002) 
\times \mathrm{FWHM}(\mathrm{H}\beta)$. The inset plot shows the distribution 
of the FWHM ratio of H$\alpha$ to H$\beta$.  Similarly, we show in Figure \ref{Fig:flux_hahb}, the flux of H$\alpha$ broad component plotted against the flux of the broad component of H$\beta$. A linear least-squares fit gives us $F(\mathrm{H}\alpha)=(3.312 \pm 0.012) 
\times F(\mathrm{H}\beta)$. These values are consistent with those existing the literature \citep[e.g. see, ZH06 and][]{2016MNRAS.457.3878Z}.  

\begin{figure*}
\centering
\resizebox{18cm}{11.0cm}{\includegraphics{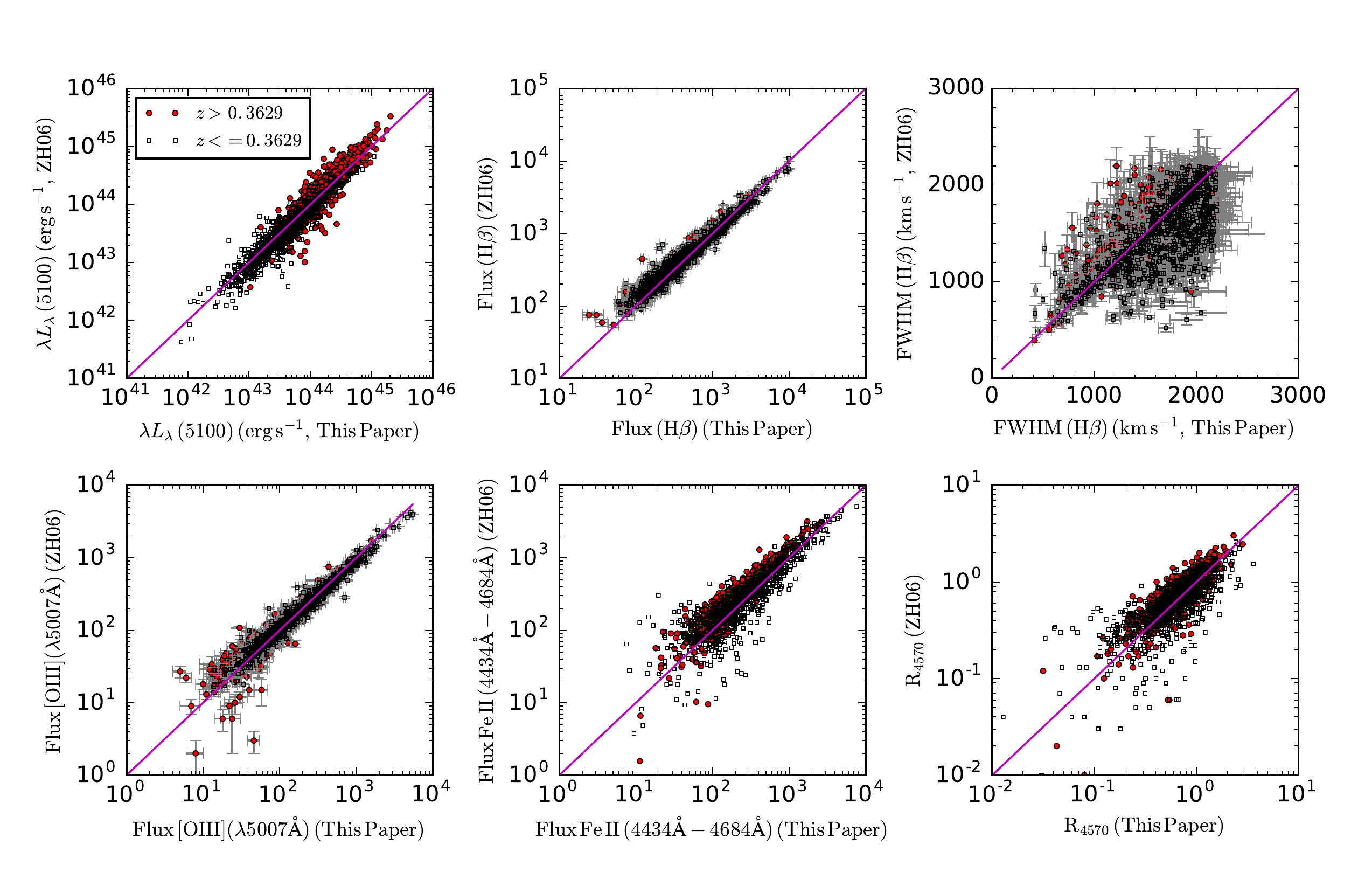}}
\caption{Comparison of our measurements with respect to ZH06 for nuclear monochromatic luminosity (upper left), broad H$\beta$ line flux (upper middle), broad H$\beta$ FWHM (upper right), flux of [O {\small III}] $\lambda 5007\,\mathrm{\AA}$ (lower left), flux of Fe II ($4434\mathrm{\AA-4684\AA}$; lower middle) and strength of Fe II (lower right). The flux values are in units of $\mathrm{10^{-17}\, erg\, s^{-1}\, cm^{-2}}$.}\label{Fig:comparison} 
\end{figure*}

\subsection{Comparison with the NLSy1 galaxies catalog of ZH06}
Using the fitting procedure outlined above, we have arrived at a new sample of NLSy1 galaxies. For a consistency check, it will also be useful to test whether in our procedure we are able to retrieve the sample of NLSy1 galaxies reported by ZH06 using SDSS DR3. For that, we compared our sample with that of ZH06. We found that 1815 out of the 2011 sources reported by ZH06 are included in our sample.   
%On comparing our sample reported here with the sample of ZH06, we found that 1815 out of the  2011 sources in ZH06  is included in our sample. 
Our fitting procedure is thus able to recover 90\% of 
the sources found by ZH06. The 10\% of sources that is in the ZH06 sample, but
missed in our catalog might be due to (1) the fact that our selection considered the
width of H${\beta}$ $\le$ 2200 km sec$^{-1}$, whereas ZH06 considered either of
the two Balmer lines\footnote{Some objects 
do not have detectable H$\beta$ but have H$\alpha$ and thus included in the sample of ZH06.} to have FWHM $\le$ 2200 km sec$^{-1}$ and (2) that the ZH06 sample has 29 NLSy1 classified as ``galaxy'' in SDSS DR3, which are not included in our parent catalog of 114,806 candidates as our selection only considers the sources classified as ``QSO'' by SDSS pipeline. For those sources that are common to ours and the ZH06 sample, we found that typical errors in the FWHM of the H$\beta$ broad component are around 5\%, which is similar to the value of 6\% quoted by ZH06. However, considering our whole sample of 11,101 sources, the median error in the FWHM of the H$\beta$ broad component is around 7\%.

In Figure \ref{Fig:comparison}, we compare our measurements with the values obtained by ZH06 for all the objects that are included in both of the catalogs. The nuclear monochromatic luminosity\footnote{Only objects having nuclear light fraction (FC) $>0.25$ in ZH06 have been plotted since the authors noted FC$\le0.25$ is unreliable.} (upper left), broad H$\beta$ line flux (upper middle), broad H$\beta$ FWHM (upper right), flux of [O {\small III}] $\lambda \, 5007 \, \mathrm{\AA}$ (lower left), flux of Fe {\small II}  ($\mathrm{4434\AA-4684\, \AA}$; lower middle), and strength of Fe {\small II}  i.e. $R_{4570}$ (lower right), defined as the ratio of Fe {\small II} flux in the wavelength range $4434\, \mathrm{\AA} – 4684\, \mathrm{\AA}$ to the H$\beta$ broad 
component flux, are shown in the plot. The circles represent objects having $z>0.3629$, i.e. for the objects in which only the H$\beta$ region is fitted while the empty squares are objects having $z<0.3629$ in which both H$\beta$ and H$\alpha$ regions are fitted simultaneously. Our estimated $\lambda L_{\lambda} (5100\mathrm{\AA})$ closely matches with the measurements of ZH06. The logarithmic ratio of these two measurements has an average of $0.02 \pm 0.17$. Similarly, close matches between measurements of H$\beta$ and [O {\small III}] line flux have also been found with the logarithmic ratio distribution having an average of $-0.04\pm 0.08$ and $-0.02\pm 0.10$ respectively. The Fe {\small II} 
flux and its strength are found to be similar to those of ZH06. The logarithmic ratio between our measurements to that of ZH06 has an average of $-0.07\pm 0.19$ and $-0.03\pm 0.17$ for Fe {\small II} flux and $R_{4570}$ respectively. The close correspondence between various physical quantities (shown in Fig. \ref{Fig:comparison}) deduced from the fitting process outlined here with those reported by ZH06 for sources that are common to the two studies, indicates the robustness of the fitting procedure adopted in this work.

The width of the H$\beta$ broad emission line we measured also closely matches with the measurement of ZH06. The ratio of these two measurements has a mean of 1.08 with a dispersion of 0.22. From the figure, it is evident that the H$\beta$ widths of some objects having $z<0.3629$, represented by empty square symbols, are overestimated in this work compare to ZH06 and deviate from the unit ratio line in some cases. This might be due to the differences in the fitting procedure adopted in this work and that used by ZH06. This overestimation of the FWHM of H$\beta$ broad component for some objects by us could lead to some genuine NLSy1 galaxies being missed out, however, it would definitely prevent non-NLSy1 galaxies from being included in the sample. Also, our ability to retrieve 90\% of the sources found by ZH06 reiterates the fact that our fitting procedure is robust, despite the minor difference in the methodology adopted in this work and that used in ZH06.

%\clearpage
\begin{center}
\begin{sidewaystable*}
\vspace*{10cm}
\caption{Emission line properties of NLSy1 Galaxies. Note. Columns are listed as follows: (1) SDSS ID (plate-mjd-fiber); (2) R.A. in degrees; (3) DEC. in degrees; (4) redshift; (5) logarithmic nuclear monochromatic luminosity at 5100 $\mathrm{\AA}$ ($\mathrm{erg\, s^{-1}}$); (6) H$\beta$ broad component flux ($\mathrm{10^{-17}\,erg \, s^{-1}\, cm^{-2}}$); (7) H$\beta$ broad component FWHM ($\mathrm{km \, s^{-1}}$);  (8) H$\beta$ narrow component flux ($\mathrm{10^{-17}\,erg \, s^{-1}\, cm^{-2}}$); (9) total (broad +narrow) [O {\small III}] $\lambda 5007\mathrm{\AA}$ flux ($\mathrm{10^{-17}\,erg \, s^{-1} \,cm^{-2}}$); (10) [O {\small III}] $\lambda 5007\mathrm{\AA}$ narrow component FWHM ($\mathrm{km \, s^{-1}}$); (11) H$\alpha$ broad component flux ($\mathrm{10^{-17}\,erg \, s^{-1} \,cm^{-2}}$); (12) H$\alpha$ broad component FWHM ($\mathrm{km \, s^{-1}}$); (13) optical Fe {\small II} strength relative to H$\beta$ broad component. The line fluxes are in the observed frame. Table 1 is published in its entirety in machine-readable format in the electronic edition. A portion is shown here for guidance regarding its form and content.} %Table 1 is published in its entirety in the electronic edition. A portion is shown here for guidance regarding its form and content.}
 \begin{minipage}{0.5\textwidth}
 	\centering {
 	\small\addtolength{\tabcolsep}{0.1pt}
 	%\resizebox{\dimexpr \linewidth+2cm\relax}{!}{%
     \begin{tabular}{ r r r r r r r r r r r r r}\hline\hline 
    ID  & RA & DEC & z & log($\lambda L_{\lambda}$) & F(H$\beta^{b}$) & FWHM (H$\beta^{bc}$) & F(H$\beta^{nc}$) & F([O {\small III}]$^{\mathrm{tot}}$) & FWHM([O {\small III}]$^{nc}$) & F(H$\alpha^{bc}$) & FWHM (H$\alpha^{bc}$) & $R_{4570}$ \\  
    %(a) & (b)   & (b)         &  (c)      &  (d) &  (e) &  ()& &  & & & \\
    (1)  & (2) & (3) & (4) & (5) & (6) & (7) & (8) & (9) & (10) & (11) & (12) & (13)\\ \hline
	1237-52762-0152	& 153.82315 	& 8.56158  	& 0.2439 & 43.00	& 70 $\pm$ 16 	& 1153 $\pm$	 215	& 0 $\pm$ 0 	&	 30 $\pm$	 3 	&	 422 $\pm$	 0  	&	 456 $\pm$	 40 	&	 1623 $\pm$	 112 	&	 0.08 \\
	2341-53738-0008	& 148.35292 	& 23.75970 	& 0.2375 & 43.74	& 147$\pm$ 9  	& 2092 $\pm$	 124	& 13$\pm$	 5 	&	 58 $\pm$	 30	&	 495 $\pm$	 93 	&	 645 $\pm$	 12 	&	 2059 $\pm$	 34  	&	 0.80 \\
	0404-51812-0255	& 30.78837  	& -0.82019 	& 0.5538 & 44.32	& 540$\pm$ 21 	& 2157 $\pm$	 72 	& 0 $\pm$	 0 	&	 73 $\pm$	 5 	&	 624 $\pm$	 0  	&	 ...  	&	 ...  	&	 0.36 \\
	0710-52203-0270	& 46.41772  	& -0.37211 	& 0.1894 & 42.63	& 48 $\pm$ 7  	& 1510 $\pm$	 264	& 6 $\pm$	 3 	&	 32 $\pm$	 7 	&	 204 $\pm$	 43 	&	 117 $\pm$	 13 	&	 1478 $\pm$	 164 	&	 0.07 \\
	4567-55589-0768	& 152.25979 	& 38.74255 	& 0.7141 & 43.47	& 42 $\pm$ 9  	& 2174 $\pm$	 405	& 1 $\pm$	 1 	&	 9  $\pm$	 6 	&	 218 $\pm$	 0  	&	 ...  	&	 ...   	&	 0.60 \\
	0401-51788-0475	& 25.29069  	& 0.10748  	& 0.4901 & 44.16	& 267$\pm$ 25 	& 1971 $\pm$	 132	& 37$\pm$	 9 	&	 61 $\pm$	 24	&	 949 $\pm$	 0  	&	 ... 	&	...   	&	 0.60 \\
	4546-55835-0798	& 14.16247  	& 8.54443  	& 0.5295 & 43.73	& 90 $\pm$ 12 	& 2003 $\pm$	 210	& 5 $\pm$	 2 	&	 20 $\pm$	 2 	&	 465 $\pm$	 0  	&	 ...	&	...  	&	 0.38 \\
	7448-56739-0145	& 144.59763 	& 50.89967 	& 0.6946 & 43.86	& 37 $\pm$ 6  	& 1324 $\pm$	 178	& 0 $\pm$ 0 	&	 23 $\pm$	 4 	&	 943 $\pm$	 0  	&	 ...  	&	 ...   	&	 0.94 \\
	2229-53823-0306	& 181.80626 	& 27.36376 	& 0.2149 & 43.53	& 221$\pm$ 13 	& 468 $\pm$	 23 	& 0 $\pm$ 0 	&	 87 $\pm$	 17	&	 323 $\pm$	 0  	&	 778 $\pm$	 47 	&	 871  $\pm$	 36  	&	 0.75 \\
	7384-56715-0868	& 153.88019 	& 46.90631 	& 0.4754 & 43.43	& 41 $\pm$ 9  	& 1925 $\pm$	 339	& 2 $\pm$	 1 	&	 17 $\pm$	 4 	&	 368 $\pm$	 0  	&	 ...  	&	 ...   	&	 0.83 \\ \hline\hline
\end{tabular} } %}
 	\label{Table:object}
 	\end{minipage}
 %\end{table*}	
 \end{sidewaystable*}
\end{center}
%Xray detected object        
%Radio detected object

%\section{Sample properties}\label{sec:properties}

%======================================================
%======================  Section ======================
%======================================================

%\section{Discussion}\label{sec:discussion}

%======================  Subsection ======================
\section{Properties of NLSy1 galaxies of Our catalog}\label{sec:discussion}
The present sample increases the number of known NLSy1 galaxies by a factor of about five. It would therefore be of interest to investigate the ensemble properties of this new sample. Such an analysis using this large sample is indeed necessary to either confirm or refute the various properties of NLSy1 galaxies that are based on a rather smaller sample size. We, therefore, present below some general characteristics of this new sample and make comparisons as and where needed. Detailed investigation of other physical properties of this sample will be presented elsewhere.

%======================  Subsubsection ======================
\subsection{Redshift and absolute magnitude}
\begin{figure}
\centering
\resizebox{9cm}{4.0cm}{\includegraphics{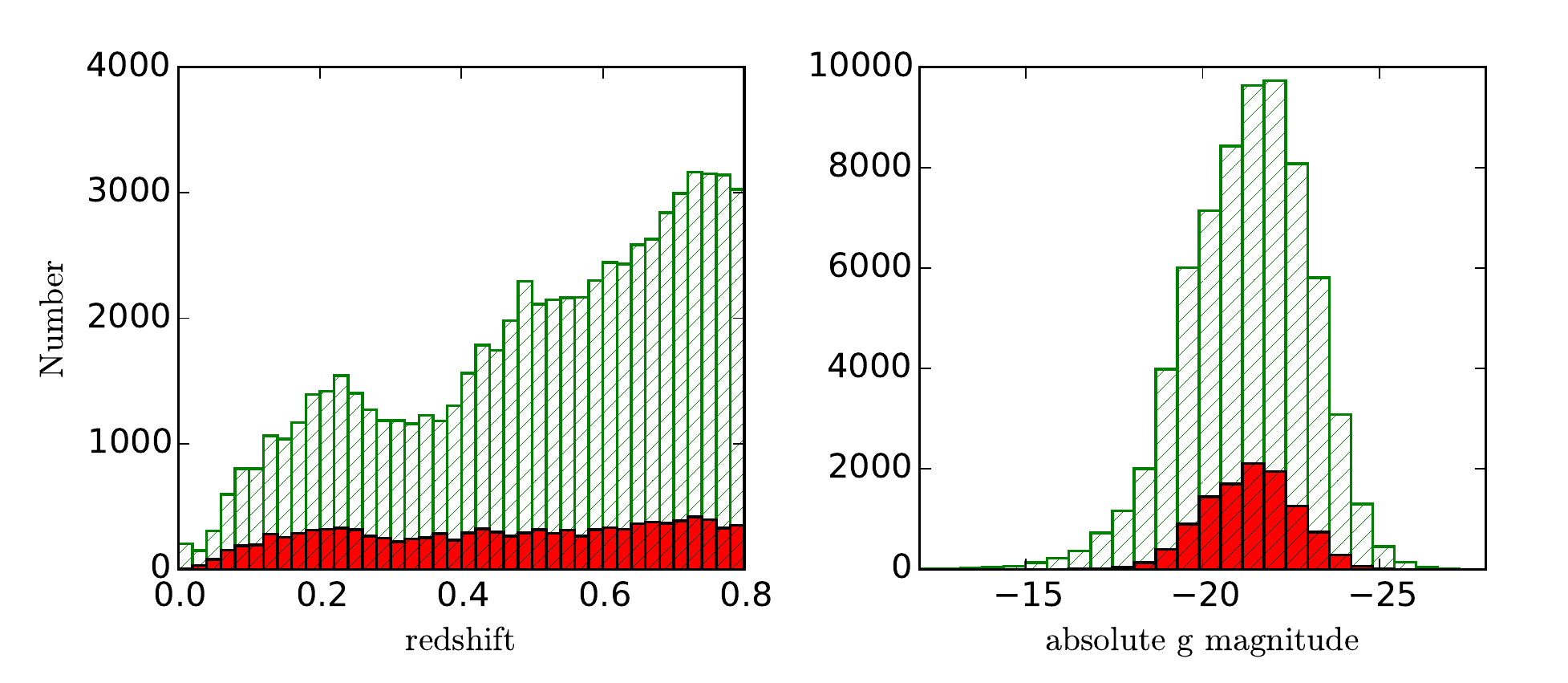}}
\caption{Distribution of redshift (left) and absolute $g$ magnitude (right) of the parent sample of BLAGN (empty-hatched) and NLSy1 galaxies (filled).}\label{Fig:redshift}. 
\end{figure}

The number of SDSS sources identified as ``QSO'' by the SDSS pipeline with $z$ 
$<$ 0.8 and a median SNR $> 2 \,\mathrm{pixel^{-1}}$ in SDSS DR12 is 68,859. From this initial set of sources, the number of NLSy1 galaxies  selected based on our criteria is 11,101, which
is about 16\% of our original sample of broad-line AGNs. This
is consistent with what is found earlier by 
\citet{1988LNP...307....1O}, \citet{2001A&A...372..730V} and ZH06. The redshift distribution and absolute $g$-band magnitude of all the BLAGN (parent 
sample: empty-hatched) and NLSy1 galaxy sample (filled region) are shown in 
Figure \ref{Fig:redshift}. The number of BLAGN increases with redshift in our 
parent sample, though the NLSy1 galaxies seem to be equally populated throughout $z<0.8$. 
On the other hand, the absolute $g$-band magnitude distribution of both
BLAGN and NLSy1 galaxies peaks at $M_g \sim -22$ mag. A similar peak in
the absolute magnitude distribution is also noted by ZH06.

\begin{figure}
\centering
\resizebox{7cm}{12.0cm}{\includegraphics{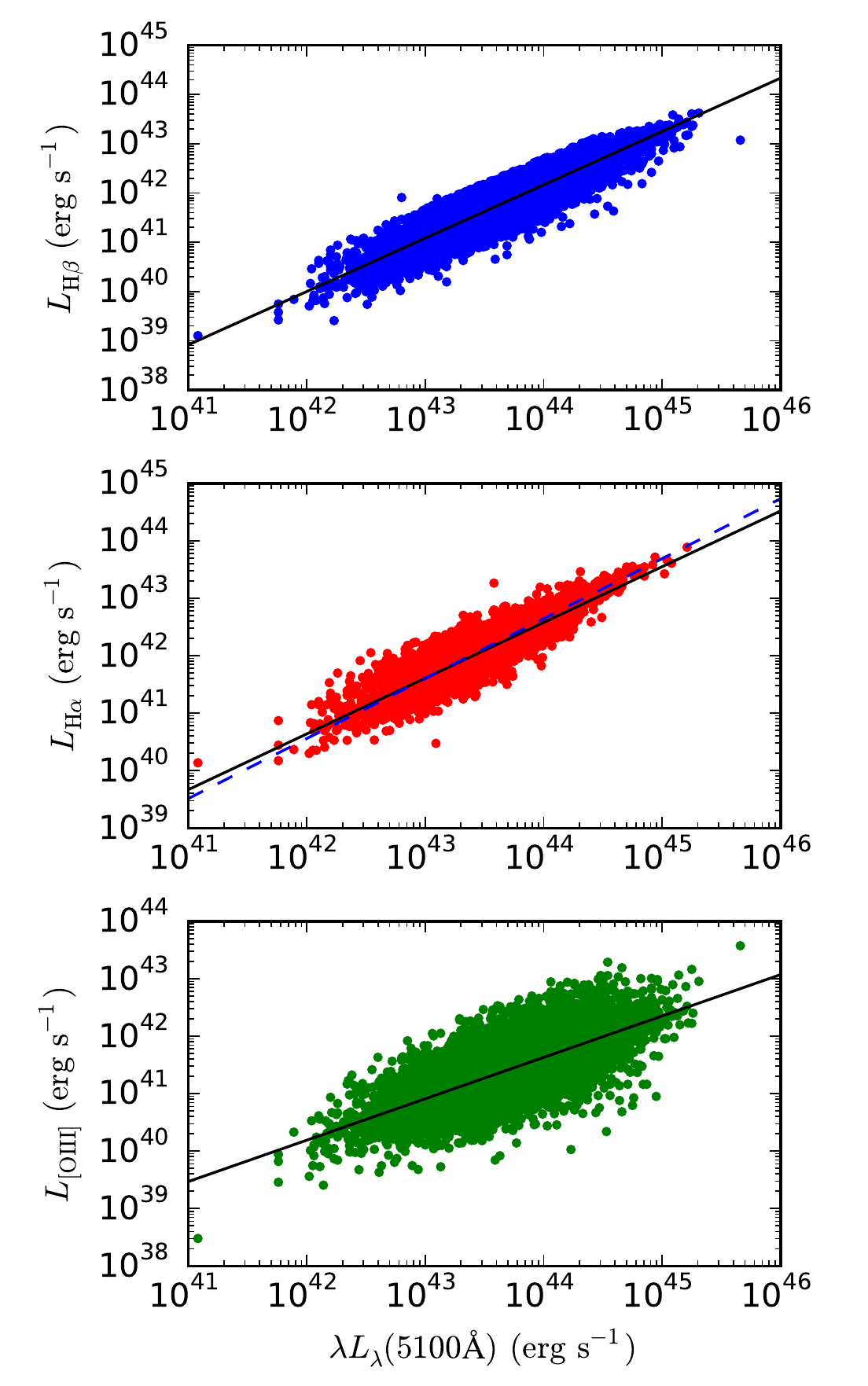}}
\caption{Correlations of the luminosity of H$\beta$ (top), H$\alpha$ (middle), and [O {\small III}] $\lambda 5007 \,\mathrm{\AA}$ (bottom) with monochromatic luminosity at 5100 $\mathrm{\AA}$. The best linear fit (solid line) is shown in each panel. The dashed line in the middle panel shows the relation of \citet{2015ApJ...806..109J}.} \label{Fig:lum}. 
\end{figure}

%======================  Subsubsection ======================

\subsection{Luminosity, equivalent width and Fe {\small II} strength}

 \citet{2005ApJ...630..122G} found a strong correlation between the luminosity of Balmer lines and optical continuum luminosity. Interestingly, such relations are valid over a wide range of luminosity ($10^{42}<L_{5100}<10^{47} \, \mathrm{erg\, s^{-1}}$) and redshift ($0<z<6$), suggesting that the response of the BLR to the incident continuum is consistent across all redshifts and luminosities i.e., the physical mechanism governing the correlation is the same in all AGNs \citep{2015ApJ...806..109J}. We estimated the luminosity of H$\beta$ and H$\alpha$ for all the objects in our sample having $0<z<0.8$ and $0<z<0.3629$ respectively. The correlation of monochromatic luminosity at 5100 $\mathrm{\AA}$ with the luminosity of H$\beta$ (top), 
H$\alpha$ (middle), and [O {\small III}] $\lambda 5007\, \mathrm{\AA}$ (bottom) is shown in Figure \ref{Fig:lum}. All correlations are very strong, as found earlier by various authors \citep[e.g.,][]{2005ApJ...630..122G,2006ApJS..166..128Z,2015ApJ...806..109J}. Linear
least-squares fit yields 
{\small 
\begin{equation}
\log(L_{\mathrm{H\beta}})=(-5.53 \pm 0.21)+ (1.084 \pm 0.004) \times \log \left(  \lambda L_{\lambda}(5100 \mathrm{\AA}) \right).
\end{equation} 
\begin{equation}
\log(L_{\mathrm{H\alpha}})= (-0.17 \pm 0.34) + (0.971 \pm 0.007) \times \log  \left( \lambda L_{\lambda}(5100 \mathrm{\AA}) \right).
\end{equation} 
\begin{equation}
\log(L_{\mathrm{[O {\small III}]}})= (9.87 \pm 0.29) + (0.721 \pm 0.006) \times \log  \left( \lambda L_{\lambda}(5100 \mathrm{\AA}) \right).
\end{equation} 
}
The dashed line in the middle panel of Figure \ref{Fig:lum} represents the relation found by \citet{2015ApJ...806..109J}, where they studied an $L_{\mathrm{H}\alpha}-L_{5100}$ relationship over a wide range of luminosity and redshift ($0<z<6$). The strong correlations between Balmer line luminosity and continuum luminosity over a wide range of luminosity and redshift suggest that the former can be used in measuring black hole mass when the latter is subjected to large uncertainty, especially in some AGNs, where the host galaxy contamination or emission from the jet is significant and optical continuum luminosity is difficult to estimate \citep[e.g.,][]{2005ApJ...630..122G}. The relation of [O {\small III}] luminosity and continuum luminosity is particularly useful to measure intrinsic luminosity and the black hole mass of narrow emission line AGNs based on [O {\small III}] luminosity  \citep[e.g.,][]{2003AJ....126.2125Z}.

\begin{figure}
\centering
\resizebox{9cm}{6.0cm}{\includegraphics{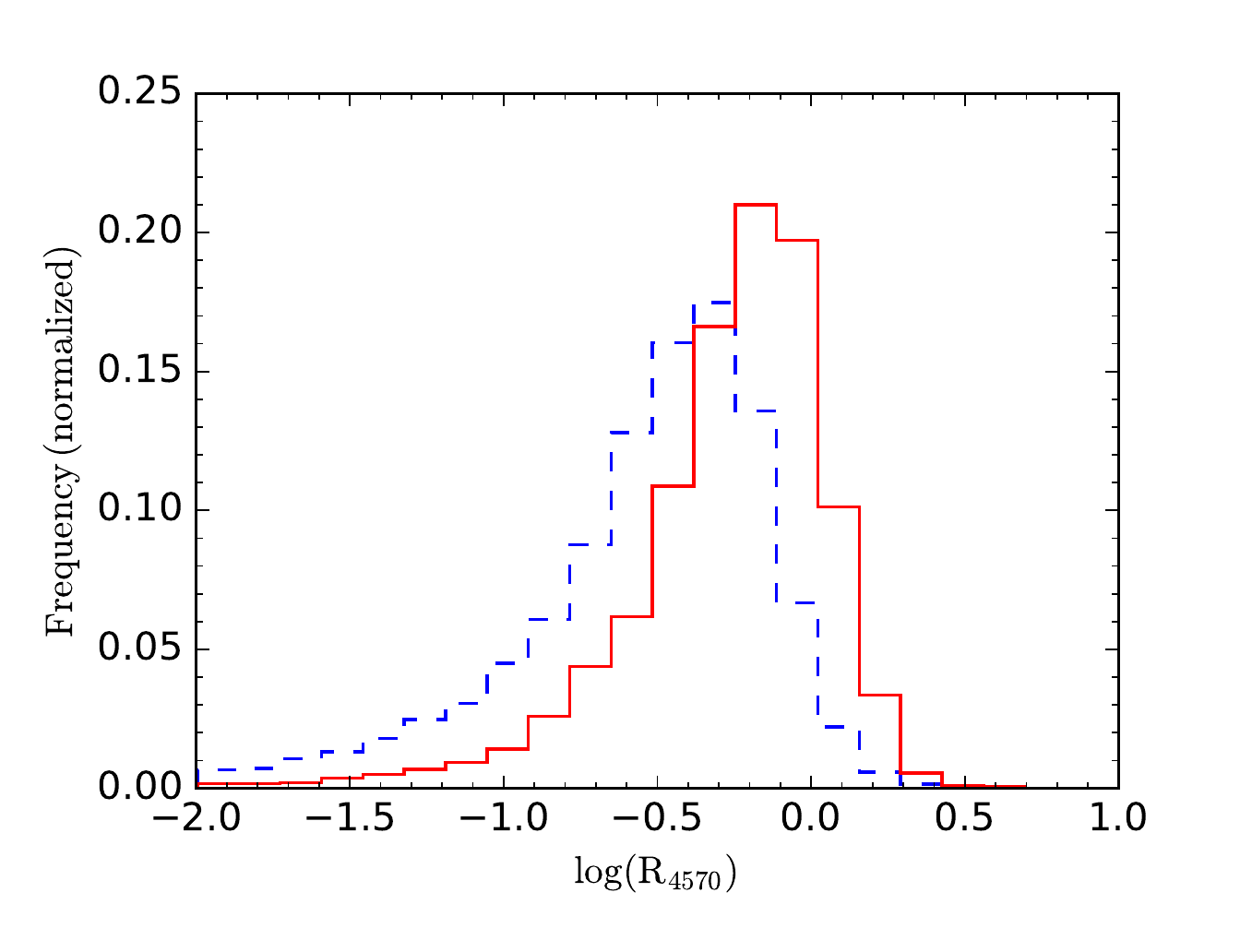}}
\caption{ Distribution of the relative strength ($R_{4570}$) of Fe {\small II} emission with respect to H$\beta$ for the NLSy1 sample (solid line) and BLSy1 sample (dashed line).}\label{Fig:R4570}. 
\end{figure}

\begin{figure}
\centering
\resizebox{9cm}{4.0cm}{\includegraphics{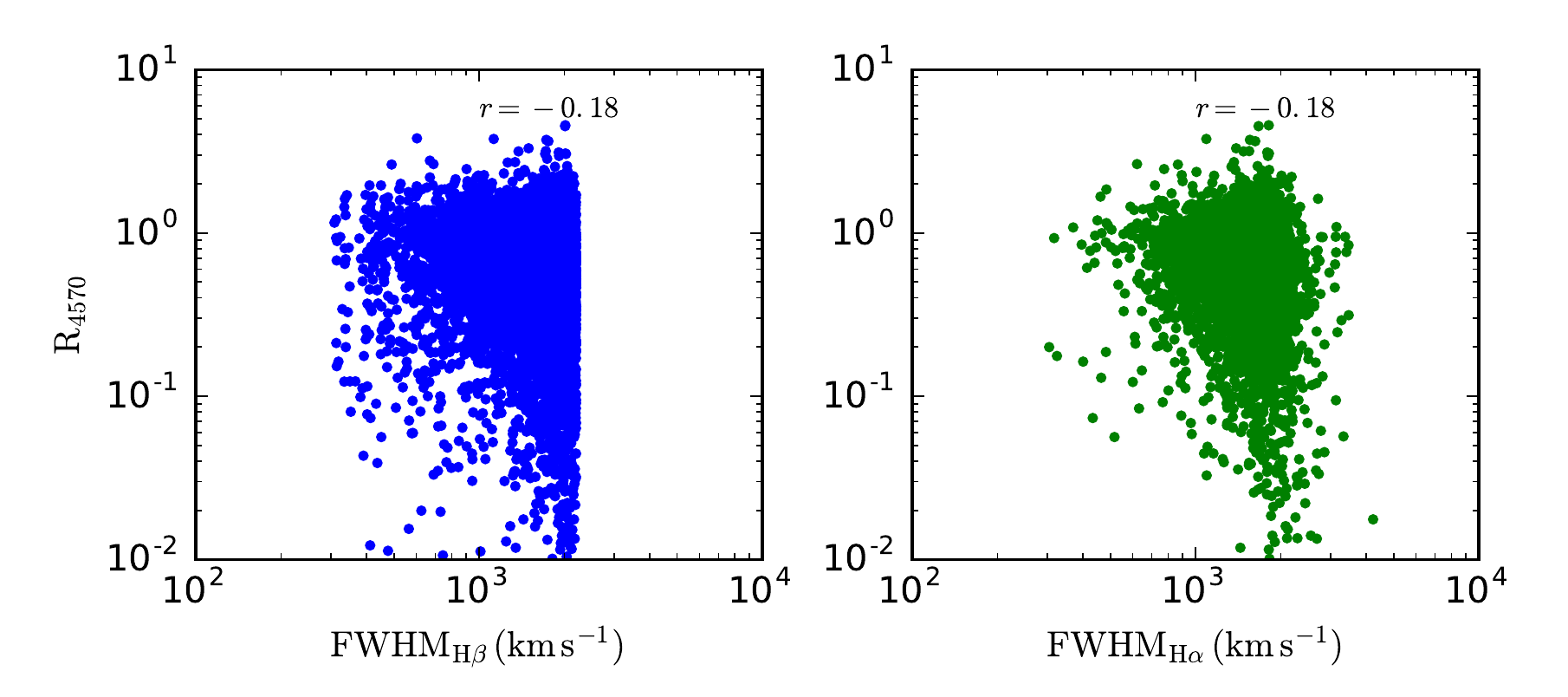}}
\caption{Correlation $R_{4570}$ with FWHM of H$\beta$ (left) and H$\alpha$ (right). The Spearman correlation coefficient ($r$) is noted in each panel.}\label{Fig:feii_width}. 
\end{figure}

One of the main characteristics of NLSy1 galaxies is that they have strong Fe {\small II} emission \citep{2001AJ....122..549V,2006ApJS..166..128Z}. The value of Fe {\small II} strength ($R_{4570}$) in a typical AGN  is $\sim 0.4$ \citep[see][and the reference 
therein]{1984MNRAS.207..263B} and about 0.8 in NLSy1 galaxies, as reported by ZH06. 
Our original sample provides us with a list of BLSy1 galaxies 
($\mathrm{FWHM_{H\beta} >2200 \, km\, s^{-1}}$), allowing us to compare the Fe {\small II} strengths between NLSy1 and BLSy1 galaxies. The distributions of 
$R_{4570}$ for both the samples are shown in Figure \ref{Fig:R4570}. The distribution for NLSy1 galaxies (solid line) peaks at larger $R_{4570}$ having a mean and 1$\sigma$ dispersion of 0.64 and 0.40, respectively, compared to the BLSy1 galaxies (dashed line), which has a mean and 1$\sigma$ dispersion of 0.38 
and 0.34 respectively. The Kolmogorov$–$Smirnov (K-S) two-sample test rejects the null hypothesis with a $p$-value of $<1\times 10^{-100}$ that the
two distributions are drawn from the same population. This is consistent with the earlier results in the literature \citep[e.g. see,][]{2001AJ....122..549V,2006ApJS..166..128Z} and clearly indicates that NLSy1 galaxies 
are stronger Fe {\small II} emitters than BLSy1 galaxies. In our sample about 60\% NLSy1 galaxies have $R_{4570}>0.5$, about 16\% 
shows moderately strong Fe {\small II} emission with $R_{4570}>1$ and about 0.5\% show 
super strong Fe {\small II} emission 
\citep[$R_{4570}>2.0$;][]{1988MNRAS.235..261L}. According to \citet{1985ApJ...291..112G}, the reason for having large 
$R_{4570}$ in NLSy1 galaxies is due to their weak H$\beta$ line instead of their strong Fe {\small II} emission.

\begin{figure}
\centering
\resizebox{8.6cm}{14cm}{\includegraphics{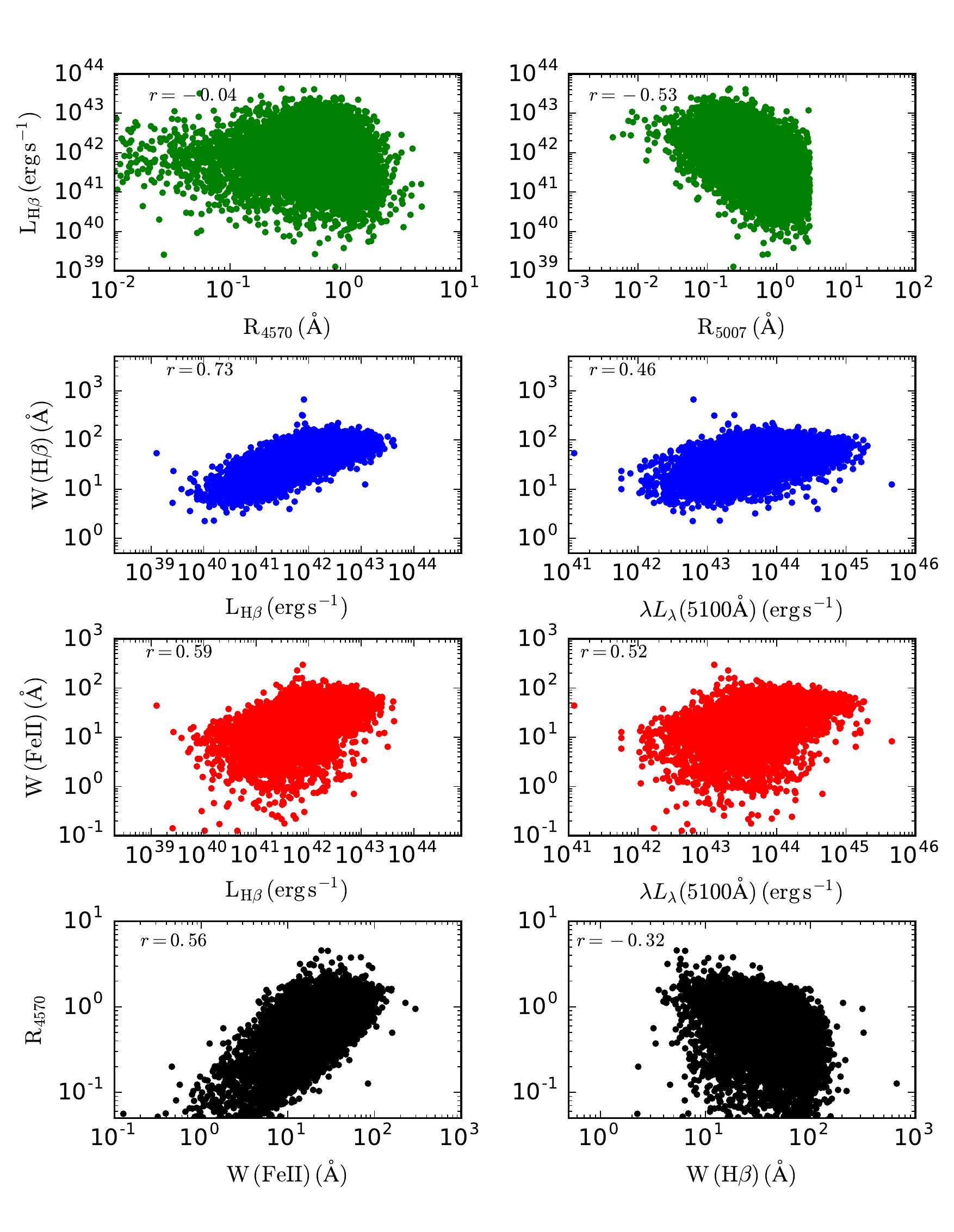}}
\caption{First panel: correlations of H$\beta$ luminosity with relative Fe {\small II} strength, $R_{4570}$, (left) and the flux ratio of [O {\small III}] ($\lambda 5007 \, \mathrm{\AA}$) to the total H$\beta$, $R_{5007}$ (right). The objects with $R_{4570} >0.01$ are only shown in the left panel and used to estimate the correlation of $L_{\mathrm{H\beta}}$-$R_{4570}$. The cut on the right panel is due to $R_{5007}<3$ used to select NLSy1. Second panel: correlation of H$\beta$ equivalent width with the luminosity of H$\beta$ (left) and monochromatic luminosity at 5100 $\mathrm{\AA}$ (right). Third panel: correlation of Fe {\small II} equivalent width with the luminosity of H$\beta$ (left) and monochromatic luminosity at 5100 $\mathrm{\AA}$ (right). Fourth panel: the correlation of $R_{\mathrm{4570}}$ with the equivalent width of Fe {\small II} (left) and H$\beta$ (right). The Spearman correlation coefficient ($r$) is noted in each panel. }\label{Fig:W_L}. 
\end{figure}

\begin{figure*}
\centering
\resizebox{18cm}{4.5cm}{\includegraphics{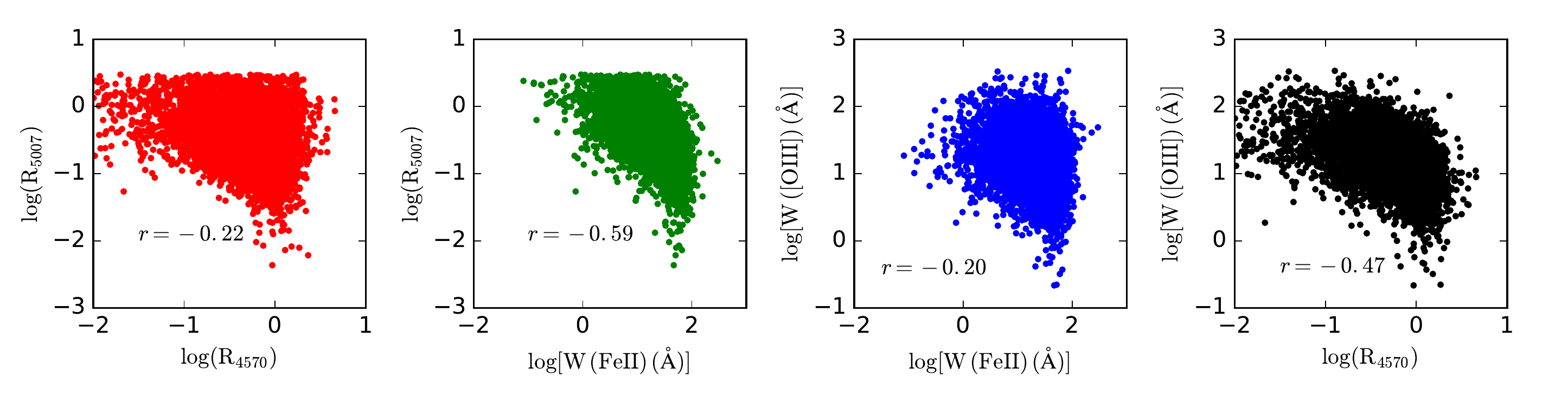}}
\caption{From left to right, the plots are $R_{4570}$ vs. $R_{5007}$, $R_{5007}$ vs. Fe {\small II} equivalent width ($W$), $W$([O {\small III}]) vs. $W$(Fe {\small II}), and $W$([O {\small III}]) vs. $R_{4570}$. The Spearman correlation coefficient ($r$) is noted in each panel. Only the objects with $R_{4570}>0.01$ are considered for correlation analysis.}\label{Fig:R5007_R4570}. 
\end{figure*}

In Figure \ref{Fig:feii_width}, we plotted Fe {\small II} strength against the  
FWHM of broad H$\beta$ (left) and H$\alpha$ (right) of our sample. It is 
clear that there is a weak anti-correlation with the Spearman correlation 
coefficient of $-0.18$ in both cases (with a $p$-value of $1\times 10^{-21}$ and $1 \times 10^{-29}$ respectively). This weak correlation in NLSy1 galaxies is consistent with the correlation found by 
ZH06. In Figure \ref{Fig:W_L} we show the correlation between different
derived parameters of our NLSy1 galaxy sample. No correlation is found
between H$\beta$ luminosity and $R_{4570}$ (top left). We also calculated the equivalent widths ($W$) of Fe {\small II} and H$\beta$ lines by the ratio of line flux to the continuum flux at 5100 $\mathrm{\AA}$ i.e., $W=F_{\mathrm{line}}/F_c(\lambda=5100\,\mathrm{\AA})$. Interestingly, both $W$(Fe {\small II}) and  $W$(H$\beta$) are found to be  strongly 
correlated with the luminosity of H$\beta$ line and monochromatic luminosity at 
5100 $\mathrm{\AA}$ (second and third panels). This is indeed similar to what was found 
by ZH06 and partially in agreement with \citet{2001A&A...372..730V} who found that the H$\beta$ luminosity is not correlated with the equivalent width of Fe {\small II} but correlated with H$\beta$. The presence of 
such strong correlations suggest that NLSy1 galaxies are indeed a different 
class of objects compared to BLSy1 galaxies because the latter generally show no or 
very weak correlation. Moreover, a strong correlation between 
$R_{\mathrm{4570}}$ and  W(Fe {\small II}) ($r=0.5$), and 
an anti-correlation between $R_{\mathrm{4570}}$ with the equivalent width of 
H$\beta$ has been found (fourth panel). This suggests that the observed large 
$R_{\mathrm{4570}}$ is not only due to weak H$\beta$ but also due to strong Fe {\small II}, which is in agreement with \citet{2016arXiv160703438C} and partially 
in disagreement with \citet{1985ApJ...291..112G}. Also shown in 
Figure \ref{Fig:W_L} is the correlation between H$\beta$ luminosity 
against $R_{5007}$, which is the ratio of total [O {\small III}] flux to total 
H$\beta$ (broad$+$narrow) flux ($R_{5007}= 
\mathrm{f[O \, {\sc III}]^{tot}/f(H\beta)^{tot}}$; the top-right corner).  
An anti-correlation is found that could be related to the weakness of 
H$\beta$ in low-luminosity objects (middle-left panel) as found 
in \citet{2001AJ....122..549V}.

Objects with strong Fe {\small II} emission are found to have weak [O {\small III}] lines 
and vice versa 
\citep{2012ApJ...744....7B,1999A&A...350..805G,1999ApJ...514...40M}; 
however, evidence of such anti-correlation has not been noted by
\citet{2001A&A...372..730V}. Several correlations of our NLSy1 galaxies are plotted in Figure \ref{Fig:R5007_R4570}. We find a weak anti-correlation 
between $R_{5007}$ and $R_{4570}$ (first panel, $r=-0.22$), a strong 
anti-correlation between $R_{5007}$ and W(Fe {\small II}) (second panel, $r=-0.59$), 
a weak anti-correlation between W([O {\small III}]) and W(Fe {\small II}) 
(third panel, $r=-0.20$), and a moderately strong anti-correlation between W([O {\small III}]) and $R_{4570}$ (fourth panel, $r=-0.47$). These observations thus confirm that Fe {\small II} strength is indeed stronger when [O {\small III}] is weak in NLSy1 galaxies.

%======================  Subsubsection ======================
\subsection{Radio Properties}
NLSy1 galaxies are generally radio quiet\footnote{The radio loudness parameter ($R$) is defined as the ratio of the radio flux density at 5 GHz to the optical flux density at 4400 $\mathrm{\AA}$ \citep{1989AJ.....98.1195K}.} with the radio loudness parameter $R<$ 10. A fraction of about 7\% 
of their population are known to be radio loud with $R>10$, and a very small 
fraction of about 2.5\% are found to be very radio-loud with $R>$ 100 \citep{2006AJ....132..531K} . The fraction of radio loud NLSy1 galaxies is indeed small 
compared to the 15\% we know in the quasar category of AGNs \citep{1989AJ.....98.1195K}. However, the radio-loud quasar fraction is found to depend on redshift and luminosity and this fraction could be larger than 20\% in low-redshift AGNs \citep{2007ApJ...656..680J}. It is important to find more radio-loud NLSy1 galaxies, due 
to the discovery of $\gamma$-ray emission in half-a-dozen of them by {\it Fermi}, 
which argues for the presence of relativistic jets in these sources \citep[e.g. see,][]{2015AJ....149...41P}.
With an aim to find more radio-loud NLSy1 galaxies, we cross-correlated
our sample with the FIRST survey (Catalog version 
14dec17\footnote{\url{http://sundog.stsci.edu/first/catalogs.html}}) within 
a search radius of 2$\arcsec$. We find that about 555 sources (5\%) are
detected by FIRST. %This is lower than that reported by ZH06, wherein, a total of 142 out of their 2011 NLSy1 galaxies are detected in FIRST.

\begin{figure}
\centering
\resizebox{7cm}{14.0cm}{\includegraphics{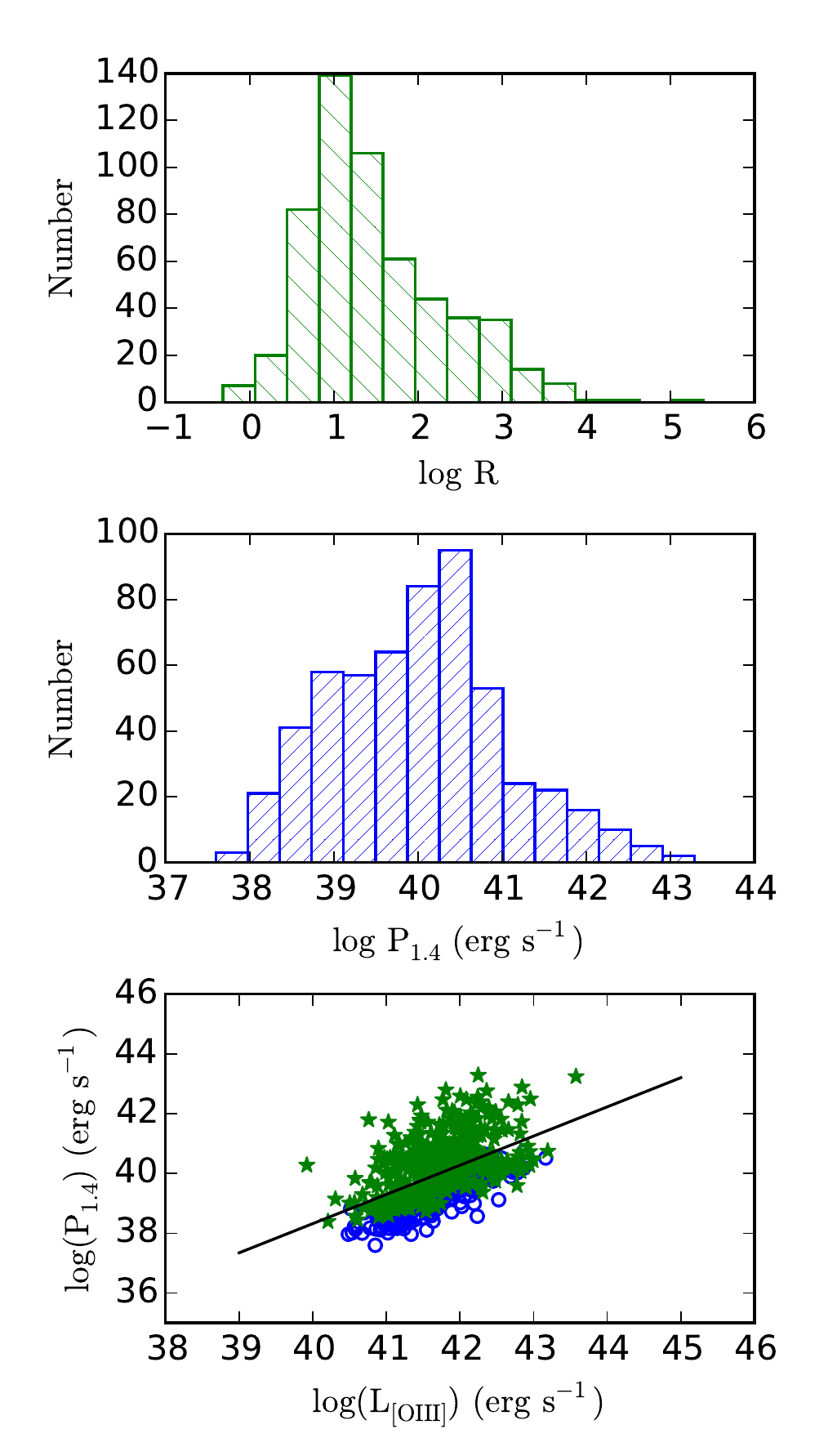}}
\caption{Distributions of logarithmic radio loudness ($\log R$, top) and radio power at 1.4 GHz (middle) of the NLSy1 in our sample. The correlation of radio power and [O {\small III}] ($\lambda 5007\, \mathrm{\AA}$) luminosity is shown at the bottom. Star symbols indicate NLSy1 with $\log\,R>1$ while empty circles indicate NLSy1 with $\log\,R<1$. The best linear fit of all radio-detected NLSy1 is shown by a solid black line. }\label{Fig:radio}. 
\end{figure}

In Figure \ref{Fig:radio}, we show the distribution of logarithmic of radio loudness (we define radio loudness as the ratio of 1.4 GHz flux to optical $g$-band flux, 
$R=\mathrm{F_{1.4\, GHz}/F_g}$) in the upper panel, the distribution of 
radio power (middle panel), and the correlation between radio power and 
luminosity of [O {\small III}] (lower panel). The radio power ($P_{1.4}$) 
was estimated using $P_{1.4}=4\pi D^2_L (1+z)^{(\beta-1)}\nu_0 F_{\nu_0}$ 
\citep{2010ApJ...720.1066C}, where $\nu_0$ is the observed frequency of 1.4 GHz, $D_L$ is the luminosity distance, and $\beta$ is the radio spectral index taken to be 0.8 \citep{1992ARA&A..30..575C}. The $\log R$ distribution has a 
mean of $1.49 \pm 0.82$. Out 
of the 555 NLSy1 galaxies that are detected by FIRST, 177 are radio 
quiet ($R<10$) and 378 are radio loud ($R>10$). Their radio power ranges between log $P_{1.4}=37.6$ to 43.3 $\mathrm{erg \,s^{-1}}$ having a peak at about 40.5 $\mathrm{erg \,s^{-1}}$. The distribution of $\log P_{1.4}$ has a mean of $40.0 \pm 1.0 \, \mathrm{erg \,s^{-1}}$, which is similar to the mean value of $\log P_{1.4}$ of $40.6 \pm 1.1 \, \mathrm{erg \,s^{-1}}$ found for the radio-detected BLSy1 galaxies in our sample indicating the influence of possible selection effects present in our sample of radio-detected sources. The 1.4 GHz radio power 
($P_{1.4}$) is found to be correlated with the luminosity of [O {\small III}]. 
The Spearman correlation coefficient $r=0.52$ with a $p$-value $1\times 10^{-41}$ 
confirms a strong positive correlation between these two parameters. A linear 
fit (solid line) yields 
\begin{equation}
\log(P_{1.4}) = (0.976\pm 0.06) \times \log \left(L_{\mathrm{[O {\small III}]}}\right) - (0.73\pm 2.78).
\end{equation} 
This is similar to the correlation between radio power at 6 cm and luminosity of [O {\small III}] found by \citet{2007ApJ...670...92G} for low-mass AGNs. However, dividing our radio-detected NLSy1 galaxies into different redshift ranges with bin size of 0.2, we found the correlation to vanish in all redshift bins except for $z<0.2$.  

% \log(P_1.4) = 0.976\pm 0.06 \log(L[O {\small III}]) -0.73\pm 2.78   

%======================  Subsubsection ======================
\subsection{$X$-$Ray$ Properties}
\begin{figure}
\centering
\resizebox{7cm}{5.0cm}{\includegraphics{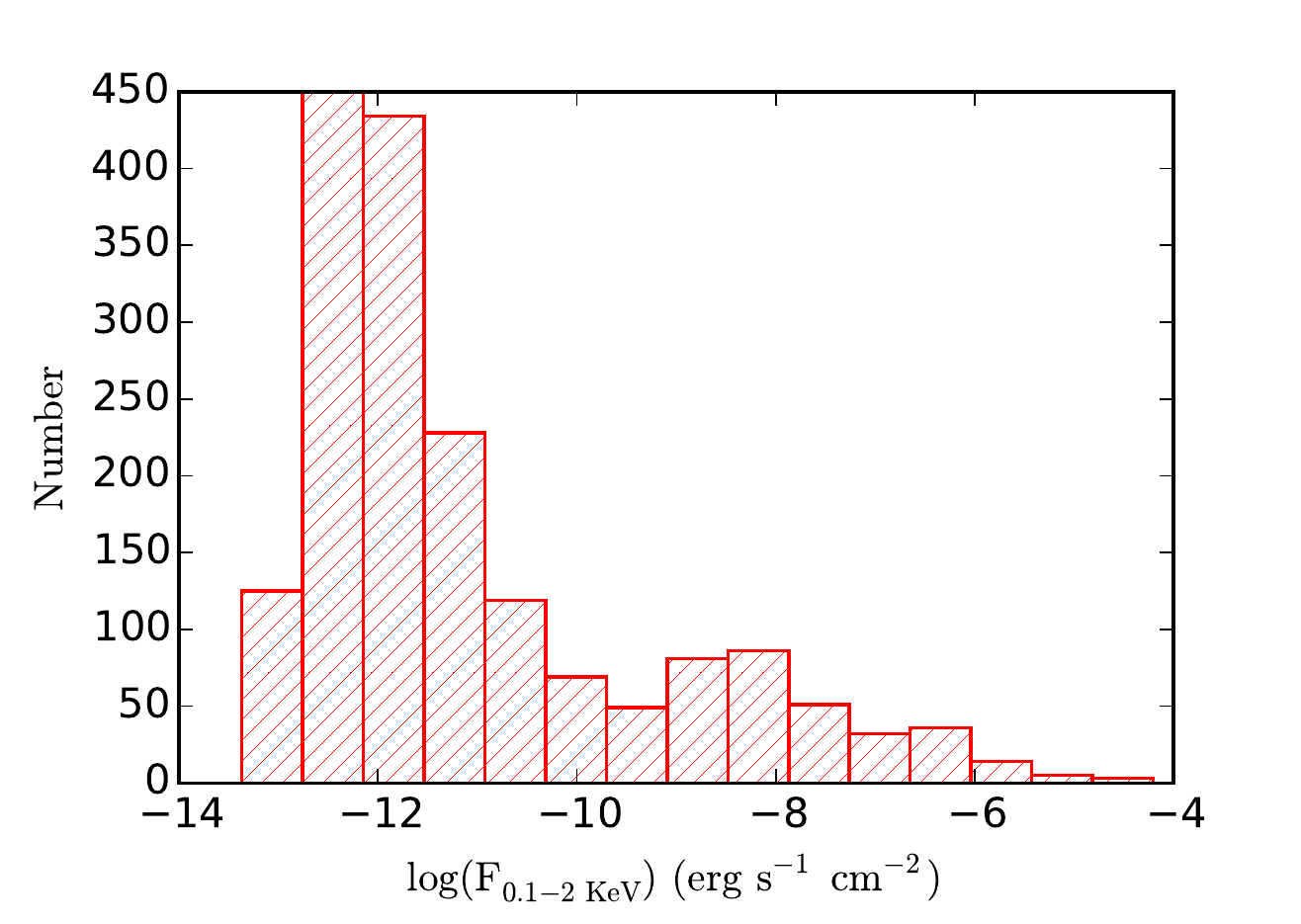}}
\caption{Distribution of soft X-ray flux of NLSy1 in our sample.}\label{Fig:X_Flux}. 
\end{figure}

NLSy1 galaxies are known to show steep soft X-ray spectra and rapid X-ray variability. 
To find the X-ray counterparts to our sample, we cross-correlated
our sample with the second \textit{ROSAT} all-sky survey (2RXS) source 
catalog \citep{2016A&A...588A.103B} within a search radius of 30$\arcsec$. This resulted in 1863 matches and amounts to 17\% 
X-ray detection. The distribution of the soft X-ray ($0.1-2$ KeV) flux of all 
those NLSy1 galaxies is shown in Figure \ref{Fig:X_Flux}. 

%======================  Subsection ======================

\subsection{Electron density of the narrow line region}
The most important density diagnostic of the NLR in AGNs is the intensity
ratio of [S {\small II}] $\lambda 6716/ \lambda 6731$ \citep{1989agna.book.....O}.
There is ambiguity in the literature on the density of [S {\small II}] emission 
region in NLSy1 vis-a-vis  BLSy1 galaxies.  Using a sample of Seyfert 1 
galaxies including seven NLSy1 galaxies, \citet{2000ApJS..126...63R} 
found that the typical density of the [S {\small II}] emission region in NLSy1 galaxies is lower than BLSy1 galaxies. According to \cite{2007ApJ...670...60X} BLSy1 galaxies avoid low-density 
($n_e<140$ cm$^{-3}$) while NLSy1 galaxies prefer low-density but show large dispersion in the density. Recently, \citet{2016arXiv160703438C} analyzed the electron density of BLSy1 and NLSy1 galaxies and found no difference in the electron-density distribution between the two classes. Because our sample of NLSy1 and BLSy1 galaxies are much larger than those used
in earlier studies, we revisited the electron-density issue of NLR between NLSy1 and 
BLSy1 galaxies. For this, we estimated the 
intensity ratio of [S {\small II}] $\lambda 6716/ \lambda 6731$ of 2551 NLSy1 and 
5533 BLSy1 galaxies. To estimate electron density from this, the python code PyNeb\footnote{\url{http://www.iac.es/proyecto/PyNeb/}}  
developed by \citet{2015A&A...573A..42L} was used. Using the PyNeb 
atomic data `IRAF\_09',  we calculated electron densities of the NLSy1 and BLSy1 galaxy sample for the intensity ratio of [S {\small II}] $\lambda 6716/ \lambda 6731$ between $0.7$ and $1.7$ with a fixed temperature $T=10^4$ K \citep[see also][]{2016arXiv160703438C}. 
The above criteria provided us density estimates of 2020 NLSy1 and 4744 BLSy1 galaxies.  

Figure \ref{Fig:ne} shows the variation of electron density 
against FWHM of H$\beta$ for both NLSy1 (dots) and BLSy1 (stars) galaxies. 
The horizontal dashed line at $n_e=140$ cm$^{-3}$ (i.e., $\log n_e=2.146$ cm$^{-3}$) separates the 
low- and high-density regions, whereas, the vertical dashed line at
2200 km s$^{-1}$ separates the sources into BLSy1 and NLSy1 galaxies. Out of the 2020 NLSy1 galaxies, 509 (25\%) have $\log n_e<2.146$ cm$^{-3}$ 
(zone A), while 1511 (75\%) have $\log n_e>2.146$ (zone B). 
In the case of BLSy1 galaxies (zone C), 725 out of 4744 objects (15\%) have $\log n_e<2.146$ 
and the remaining 4018 objects (85\%) have $\log n_e>2.146$ cm$^{-3}$. Our study suggests that about 15\% of BLSy1 galaxies in our sample have $\log n_e<2.146$ cm$^{-3}$, which is not in agreement with \citet{2007ApJ...670...60X} who found BLSy1 galaxies to have $\log n_e>2.146$ cm$^{-3}$. This disagreement is  likely due to the low number statistics of objects used by \citet{2007ApJ...670...60X} in their analysis. However, we find a weak trend of 
low-density NLR in NLSy1 and a high-density NLR in BLSy1 galaxies. In fact, a very weak but positive correlation between $n_e$ and FWHM(H$\beta$) is found using the Spearman test of correlation having a correlation coefficient of $r=0.12$, and a $p$-value of $1\times 10^{-23}$ suggesting lower electron density in the NLR of NLSy1 galaxies than BLSy1 galaxies. The density distributions for NLSy1 (solid) and BLSy1 (dashed) galaxies is shown in the inset of  Figure \ref{Fig:ne}.  We find mean $\log{n_e}$ values  of 
2.40 $\pm$ 0.47 cm$^{-3}$ and 2.54 $\pm$ 0.44 cm$^{-3}$ 
for NLSy1 and BLSy1 galaxies respectively. {Applying the K-S statistic, we 
obtained $r=0.148$ and $p$-value of $7\times 10^{-28}$ rejecting 
the null hypothesis that both the distributions are drawn from 
the same population.        
 
\begin{figure}
\centering
\resizebox{8.5cm}{6.0cm}{\includegraphics{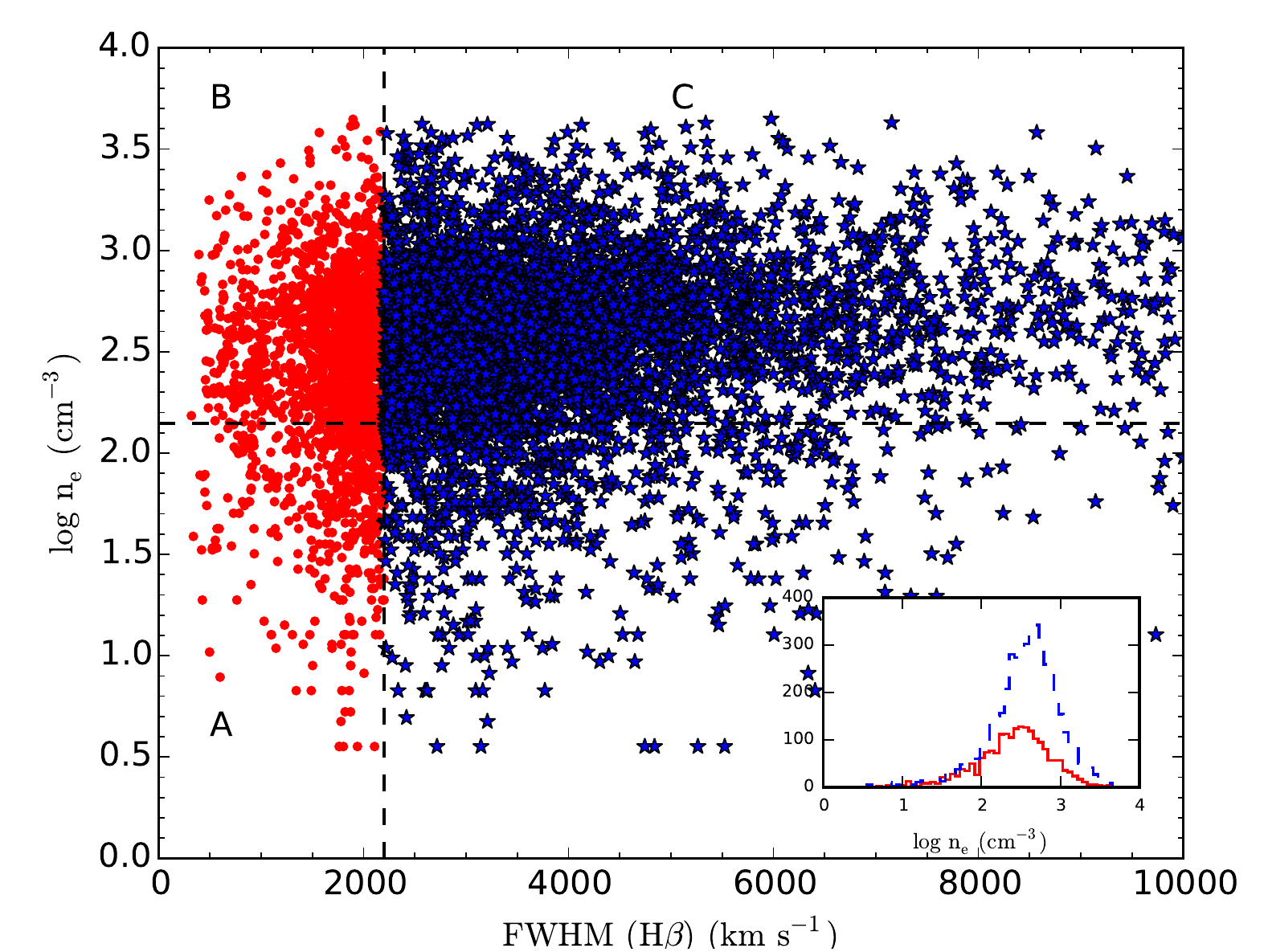}}
%\put(-60,22){{\tiny $\log \mathrm{n_e}$}}
\caption{Relation between electron density and FWHM of H$\beta$. The dots and stars represent the density of NLSy1 and BLSy1 respectively. The dashed horizontal and vertical lines indicate density 140 $\mathrm{cm}^{-3}$ and the dividing line width (i.e. FWHM(H$\beta$)=2200 $\mathrm{km\, s^{-1}}$) between two samples respectively. The lower inset plot shows a density histogram for NLSy1 (solid line) and BLSy1 (dashed line).}\label{Fig:ne}. 
\end{figure}

\subsection{Black hole mass and Eddington ratio}
The mass of the central super massive black hole of our sample of NLSy1 galaxies, as well as the BLSy1 galaxies, were estimated using the virial relation. Considering virial motion of BLR clouds, black hole mass ($M_{\mathrm{BH}}$) can be written as 
\begin{equation}
M_{\mathrm{BH}}=fR_{\mathrm{BLR}}\Delta v^2/G,
\end{equation}
where $\Delta v$ is the FWHM of broad emission line and $R_{\mathrm{BLR}}$ is the radius of the BLR \citep{1999ApJ...526..579W,2000ApJ...533..631K}. The factor $f$ known as scale factor depends strongly on the geometry and kinematics of the BLR \citep{2015MNRAS.447.2420R}. Considering the spherical distribution of clouds, we used $f=3/4$ and estimated 
$R_{\mathrm{BLR}}$ from reverberation mapping scaling relation, 
\begin{equation}
\mathrm{log}(\frac{\mathrm{R_{\mathrm{BLR}}}}{\mathrm{lt-day}}) =  K\, + \,\alpha \times \mathrm{log} \left( \frac{\lambda L_{\lambda} (5100 \mathrm{\AA})}{10^{44}}\, \mathrm{erg\, s^{-1}} \right),
\end{equation}
where $K$ (1.527) and $\alpha$ (0.533) are taken from \citet{2013ApJ...767..149B} who 
carefully subtracted the host galaxy contribution to calibrate the 
relationship between BLR size and the monochromatic luminosity at 5100 $\mathrm{\AA}$. 
The Eddington ratios $\xi_{\mathrm{Edd}}$, defined as 
$L_{\mathrm{bol}}/L_{\mathrm{Edd}}$ were estimated considering $L_{\mathrm{Edd}}=1.3\times 10^{38} M_{\mathrm{BH}}/M_{\odot}\, \mathrm{erg \, s^{-1}}$. Following \citet{2000ApJ...533..631K}, we estimated $L_{\mathrm{bol}}$ assuming $L_{\mathrm{bol}}=9\times \lambda L_{\lambda}(5100\mathrm{\AA}) \, \mathrm{erg\,s^{-1}}$ \citep[see also][]{2012AJ....143...83X}.    

The distribution of black hole mass is plotted in the left panel of 
Figure \ref{Fig:Mass} for both NLSy1 (solid line) and BLSy1 (dashed line) galaxies. In the case of NLSy1 galaxies, the $\log(M_{\mathrm{BH}})$ distribution has a 
mean of 6.9 $M_{\odot}$ with a dispersion of 0.41 $M_{\odot}$, while in the 
case of BLSy1 galaxies, the $\log(M_{\mathrm{BH}})$ distribution has a mean of 8.0 $M_{\odot}$ with a dispersion 
of 0.46 $M_{\odot}$. A K-S test applied on the 
distributions points that both distributions are remarkably 
different having a K-S statistic value of 0.83 and a $p$-value of $<1\times 10^{-150}$. NLSy1 galaxies thus have lower $M_{\mathrm{BH}}$ than BLSy1 galaxies. The right panel of Figure \ref{Fig:Mass} shows the distribution of the Eddington ratio. The log$(\xi_{\mathrm{Edd}})$ distribution for the sample of NLSy1 and
BLSy1 galaxies has mean values of $-0.34$ and $-1.03$ with standard deviations of $0.34$ and $0.42$ respectively. This suggests that NLSy1 galaxies have higher
Eddington ratios than BLSy1 galaxies. A K-S test confirms the significant 
difference between the two Eddington ratio distributions with a K-S statistic 
of 0.64 and a $p$ value $<1\times 10^{-150}$. Thus, our results are in agreement
with those available in the literature that NLSy1 galaxies have lower black hole mass and 
higher Eddington ratio than BLSy1 galaxies \citep[e.g.][]{2012AJ....143...83X}.

\begin{figure}
\centering
\resizebox{9cm}{4.0cm}{\includegraphics{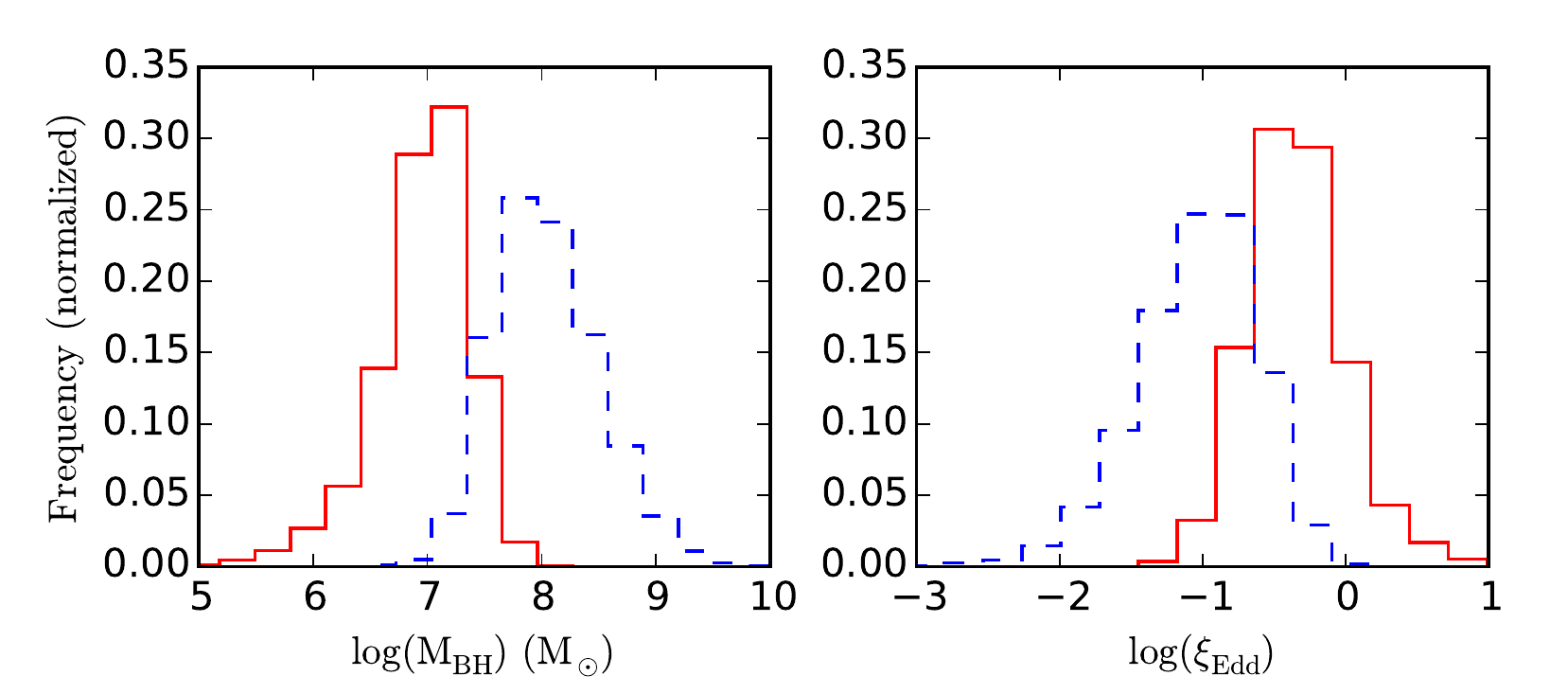}}
\caption{Distributions of black hole mass (left) and Eddington ratio (right). The NLSy1 is shown by a solid line while BLSy1 is shown by a dashed line. NLSy1 has lower mass and higher Eddington ratio than BLSy1.}\label{Fig:Mass}.
\end{figure}

\subsection{$M_{\mathrm{BH}}-\sigma_{*}$ relation}
To understand the evolution and growth of the black hole it is important to know the connections, if any, between $M_{\mathrm{BH}}$ and host galaxy properties. Available observational evidence points to a close correlation between
 $M_{\mathrm{BH}}$ and bulge mass \citep{1995ARA&A..33..581K,
1998AJ....115.2285M} as well as $M_{\mathrm{BH}}$ and $\sigma_*$ 
\citep{2000ApJ...539L...9F,2000ApJ...539L..13G,2013ARA&A..51..511K,2013ApJ...764..184M}, which suggests stellar velocity as a fundamental parameter of black hole evolution. 
Seyfert galaxies are also known to follow the same $M_{\mathrm{BH}}-\sigma_*$ 
as that shown by the normal galaxy population \citep{2004ApJ...615..652N,2010ApJ...716..269W,2013ApJ...772...49W}. Since 
NLSy1 galaxies have low-mass black holes, it is crucial to test the extension of the $M_{\mathrm{BH}}-\sigma_*$ relation to the low-mass end of NLSy1 galaxies \citep{2001NewA....6..321M}, which is attempted here for our sample. For this, we considered only those sources 
for which the spectra have a median SNR $>10 \, \mathrm{pixel^{-1}}$, and obtained $\sigma_{*}$ for 1789 NLSy1 galaxies. The $\sigma_{*}$ is the width (second-order moment) of the best-fitted Gaussian broadening function used in the convolution of our SSP template (see Eq. \ref{eq:SSP}). In Figure \ref{Fig:m-sigma}, the top panel shows the 
variation of viral black hole mass against $\sigma_{*}$. The solid line is the best linear least-squares fit to the data. A moderately strong 
correlation is found between the two parameters with a Spearman rank 
correlation coefficient 0.48 and a $p$-value of $7\times 10^{-103}$. Using the well-known $M_{\mathrm{BH}}-\sigma_{*}$ 
relation for inactive galaxies as defined by \citet{2002ApJ...574..740T}, 
\begin{equation}
\log\left( \frac{M_{\mathrm{BH}}}{M_{\odot}} \right)=8.13 + 4.02 \times \log \left( \frac{\sigma_{*}}{200\, \mathrm{km\, s^{-1}}} \right),
\end{equation}
we calculated the $M_{\mathrm{BH}, \sigma_{*}}$ using $\sigma_{*}$. This 
relation is indicated by a dashed line in the figure. About 50\% of the 
NLSy1 galaxies of our sample lie below the dashed line and $M_{\mathrm{BH}}$
values at the low-mass end have large error bars. This suggests that NLSy1 galaxies
do not follow the $M_{\mathrm{BH}}-\sigma_{*}$ relation of inactive galaxies. This closely follows the finding by ZH06, however, it is in contrast to that of \citet{2015ApJ...801...38W} who report the NLSy1 galaxies follow the $M_{\mathrm{BH}}-\sigma_{*}$ relation of BLSy1 and inactive galaxies. 

%The $M_{\mathrm{BH}}$ masses obtained for a few NLSy1 galaxies by modeling the observed optical/UV fluxes using a Shakura Sunyaev disk by \citet{2013MNRAS.431..210C} is found to be larger than the virial $M_{\mathrm{BH}}$ estimated here. If it is indeed true for the population of NLSy1 galaxies as a whole, it is likely that NLSy1 galaxies too (similar to BLSy1 galaxies) obey the $M_{\mathrm{BH}}- \sigma_{*}$ relation of inactive galaxies.

\begin{figure}
\centering
\resizebox{8cm}{6.0cm}{\includegraphics{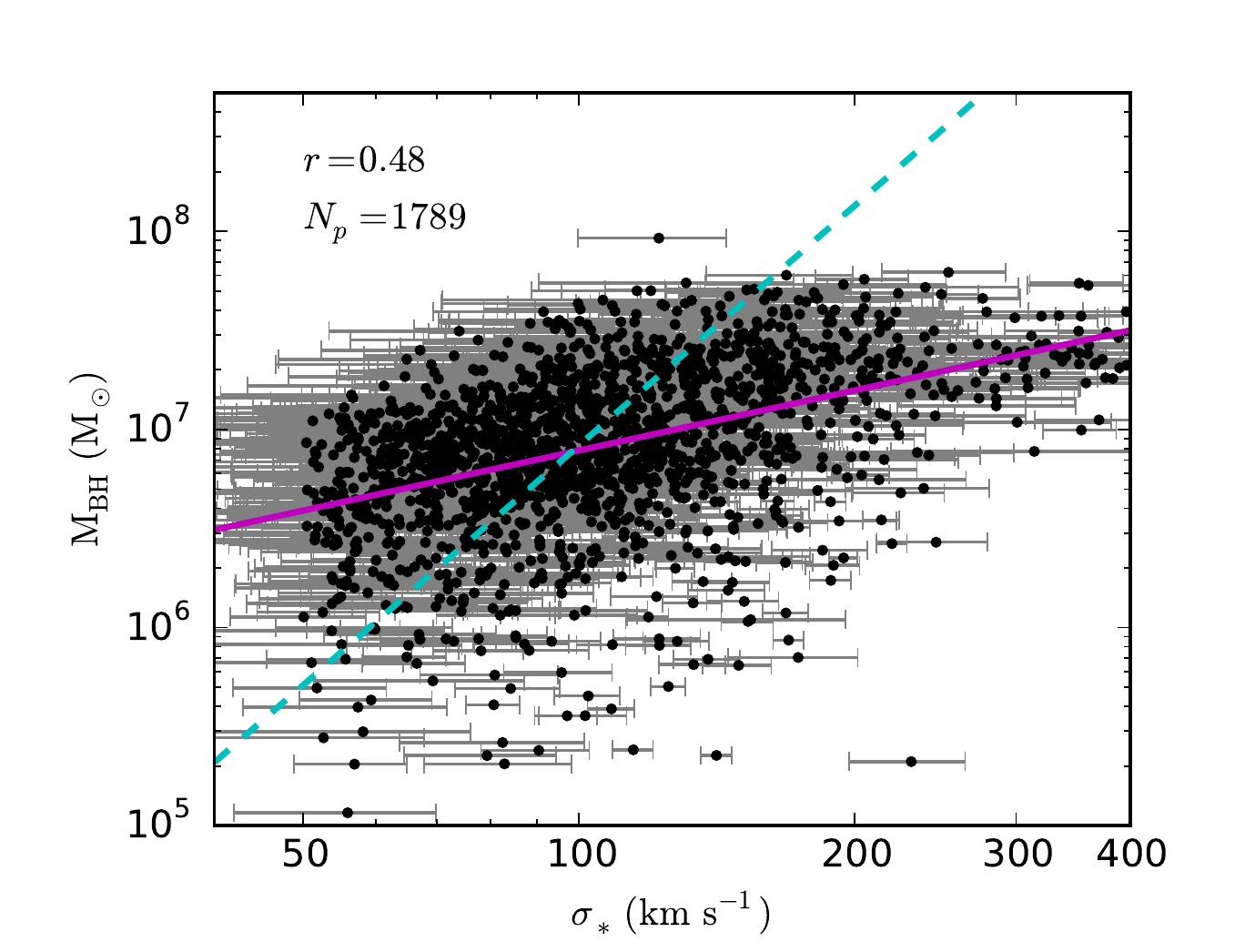}}
\caption{Correlation of black hole mass with host galaxy velocity dispersion $\sigma_*$. The dashed line shows the expected relation of \citet{2002ApJ...574..740T} and the solid line shows the best linear fit. The label $N_p$ indicates the total number of NLSy1 galaxies used in the plot.}\label{Fig:m-sigma}. 
\end{figure}

\citet{2011ApJ...739...28X} presented accurate measurements of $\sigma_*$ for 76 Seyfert 1 galaxies with low black hole mass using high-resolution spectra obtained from Keck Echellette Spectrograph and Imager (ESI) and the Magellan Echellette (MagE). Out of these, 22 objects are present in our catalog. However, only for 12 objects, we have a $\sigma_*$ measurement, non-zero at 3$\sigma$ level. The $\sigma_*$ values of those objects are plotted in Figure \ref{Fig:m-sigma_xiao}. The solid line represents the unit ratio line. Note that the spectra of all these 12 objects are obtained by SDSS with a fiber diameter of 3$\arcsec$. The ratio of $\sigma_*$ between our measurement and that of the \citet{2011ApJ...739...28X} is $0.80\pm0.35$. The large scatter might be due to low resolution as well as poor S/N in the spectra. However, the figure indicates that our result largely follows the result of \citet{2011ApJ...739...28X} indicating that SDSS spectra could be used to study $\sigma_*$ for a large sample. However, to accurately estimate $\sigma_*$ high-resolution spectra with good S/N is needed where detailed modeling of the host galaxy contribution can be carried out.

\begin{figure}
\centering
\resizebox{8cm}{7.0cm}{\includegraphics{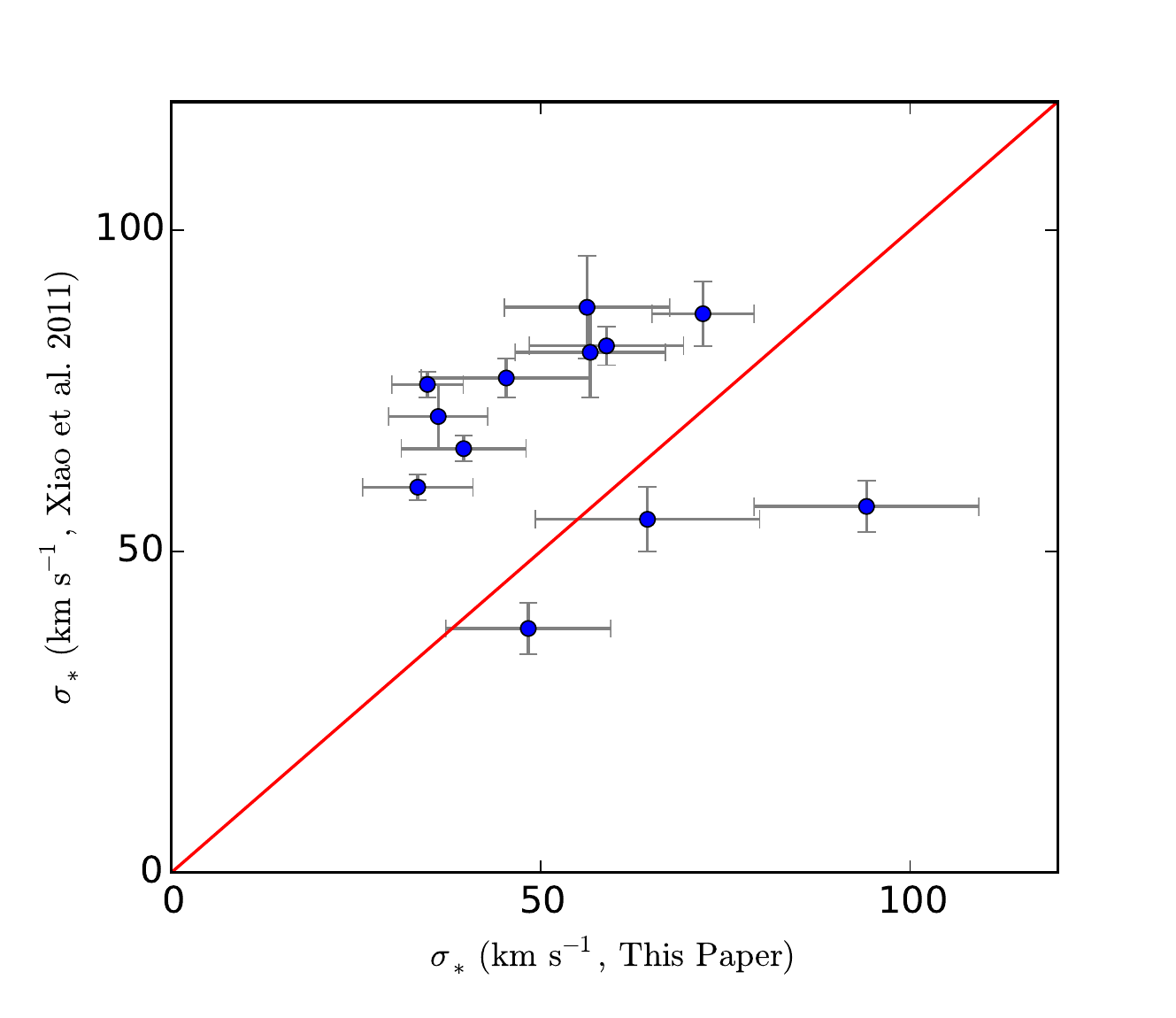}}
\caption{Values of $\sigma_*$ as obtained in this paper is plotted along the x-axis and the values obtained by \citet{2011ApJ...739...28X} is plotted along the y-axis. The unit ratio line is represented by a solid line.} %Bottom: Distribution of the $\sigma_*$/$\mathrm{FWHM([O {\small III}]_n)}$ around the best fitted value.  The best fitted Gaussian has been plotted in solid line having a 1$\sigma$ error of 1.25.}
\label{Fig:m-sigma_xiao}. 
\end{figure}

\begin{figure}
\centering
\resizebox{8cm}{12.0cm}{\includegraphics{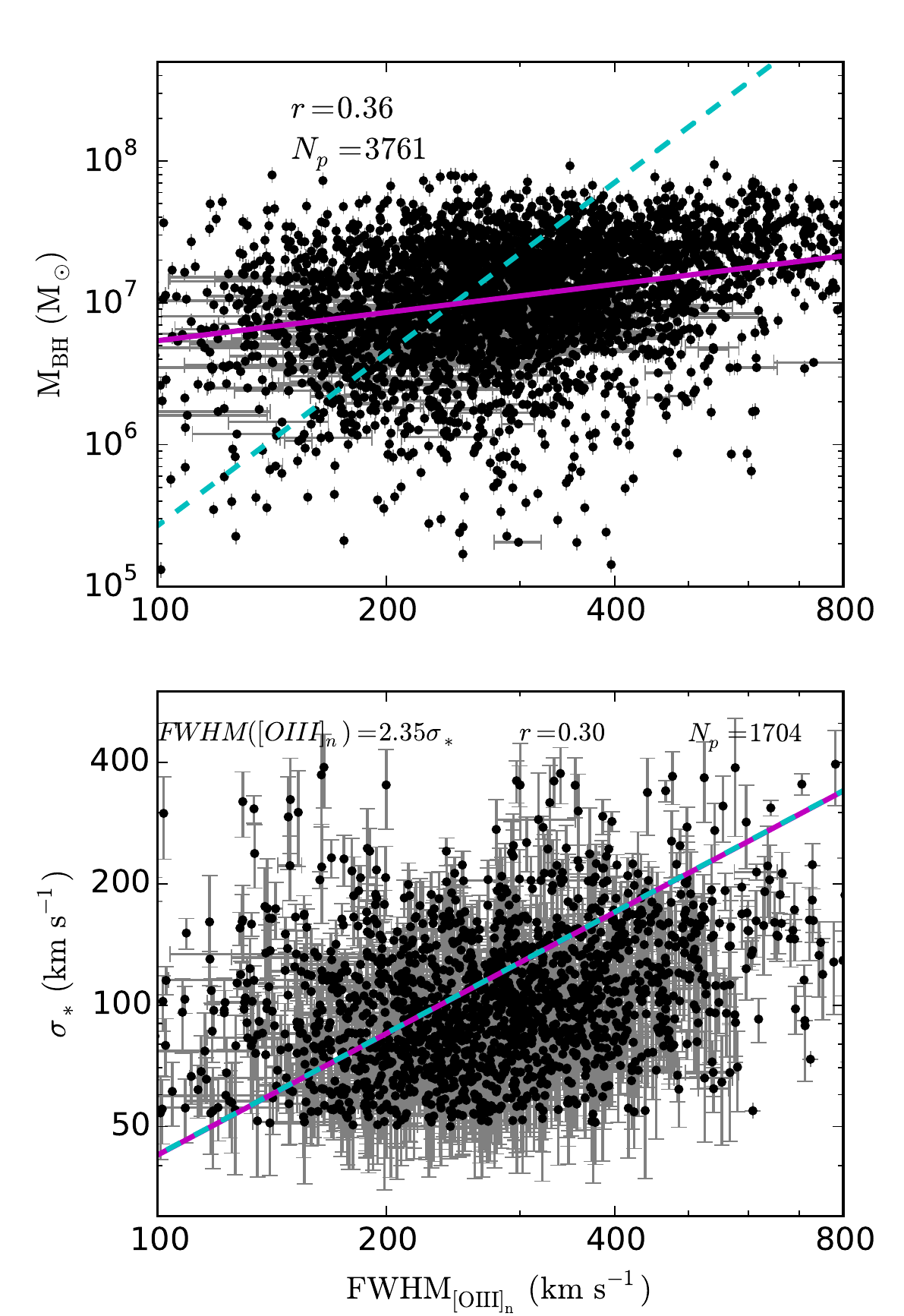}}
\caption{Top: correlation of black hole mass with $\mathrm{FWHM([O {\small III}]_n)}$. The dashed line shows the expected relation of \citet{2002ApJ...574..740T}, considering $\sigma_*=\mathrm{FWHM([O {\small III}]_n)}/2.35$, and the solid line shows the best linear fit. Bottom: plot of $\sigma_*$ against $\mathrm{FWHM([O {\small III}]_n)}$. The best linear fit to the relation is shown by a solid line having a slope $\mathrm{FWHM([O {\small III}]_n)}$=2.35$\sigma_*$, which matches the dashed line of the FWHM$=2.35\times$standard deviation.} %Bottom: Distribution of the $\sigma_*$/$\mathrm{FWHM([O {\small III}]_n)}$ around the best fitted value.  The best fitted Gaussian has been plotted in solid line having a 1$\sigma$ error of 1.25.}
\label{Fig:m-sigma2}. 
\end{figure}

Measuring $\sigma_*$ of high-redshift Seyfert galaxies is very difficult  
mainly because their cores outshine the host. However, the narrow component of 
[O {\small III}] line was used as a surrogate of $\sigma_*$, though using it 
various authors reported conflicting results. Most of these studies found that 
NLSy1 galaxies fall below the relation than the BLSy1 galaxies \citep{2001NewA....6..321M,2004AJ....127.3168B,2004ApJ...606L..41G}. On the contrary, 
\citet{2001A&A...377...52W} and \citet{2002ApJ...565..762W} found that most of the 
NLSy1 galaxies fall on the line. The origin of such a conflicting result could be due to the asymmetry in [O {\small III}] line \citep{2001A&A...372..730V,2016arXiv160703438C} because of which \citet{2001A&A...377...52W} used the 
width of [O {\small III}] line after subtracting the blue wings. A similar 
argument was also put by \citet{2007ApJ...667L..33K} who found that the width 
of the [O {\small III}] line can be used only after subtracting blue wings and 
excluding the core component of [O {\small III}] lines with strong blueshifts. 

To revisit the use of the width of the [O {\small III}] line as a surrogate of $\sigma_{*}$, we estimated the width of the narrow component of the [O {\small III}] 
line ($\mathrm{FWHM_{[O {\small III}]_n}}$) for all the NLSy1 galaxies, but use only those with a median $\mathrm{S/N}>10\, \mathrm{pixel^{-1}}$ and the narrow component of [O {\small III}] detected with more than 3$\sigma$. 
We further deconvolved the lines to remove the effect of SDSS spectral 
resolution and plotted $M_{\mathrm{BH}}$ against 
$\mathrm{FWHM_{[O {\small III}]_n}}$ in the upper panel of Figure \ref{Fig:m-sigma2}. A correlation is indeed present though weak mainly due to large scatter in the latter. Furthermore, in the lower panel of Figure \ref{Fig:m-sigma2}, we plotted $\sigma_*$ against $\mathrm{FWHM_{[O {\small III}]_n}}$. The best linear fit indicated by the solid line, $\mathrm{FWHM_{[O {\small III}]_n}}=2.35 \sigma_{*}$, is superimposed on the dashed line of FWHM$=2.35\times$standard deviation. There is an indication that many of the objects are falling off the line but large scatter in the width prevents us from making any firm statement. However, the plot clearly indicates that $\mathrm{FWHM_{[O {\small III}]_n}}$ can be a proxy of $\sigma_{*}$ as previously claimed and widely used in various literature as mentioned above. %Further, we also note that the relationship could be stronger after taking into account the suggestions made by \citet{2007ApJ...667L..33K} on the use [O {\small III}] as a surrogate of $\sigma_{*}$. 

%Note that we plotted all the objects for which narrow component of [O {\small III}] has been detected with 3$\sigma$ and spectra having $SNR>10$ a[O {\small III}]   
%We also find that the width of [O iii] l5007 is a good surrogate for j∗, but only after (1) removal of asymmetric blue wings and, more importantly, after (2) excluding core [O iii] lines with strong blueshifts (i.e., excluding galaxies which have their [O iii] velocity fields dominated by radial motions, presumably outflows). 

\subsection{Effect of inclination}
The reasons behind small $M_{\mathrm{BH}}$ and large Eddington ratios in 
NLSy1 galaxies is still unclear. The small Balmer line width that gives
rise to small $M_{\mathrm{BH}}$ based on the virial relation could be due to geometrical effects. Increasing evidence suggests that NLSy1 galaxies can have $M_{\mathrm{BH}}$ and Eddington ratios similar to BLSy1 galaxies. For example, a recent spectro-polarimetric study of the radio-loud NLSy1 galaxy PKS 2004$-$447 by \citet{2016MNRAS.458L..69B} revealed that the 
width of the H$\alpha$ line in the polarized spectrum is 9000 km/s, which is six
times broader than the width seen in direct light, yielding a $M_{\mathrm{BH}}$ of $6\times 10^8 M_{\odot}$ and much higher than the $5\times 10^{6} M_{\odot}$ 
estimated using the H$\beta$ line seen in direct light. Also, 
\citet{2013MNRAS.431..210C}, by modeling the 
optical and UV data for a sample of 23 radio-loud NLSy1 galaxies 
with a  Shakura \& Sunyaev disk, found $M_{\mathrm{BH}}$ larger than  $10^{8} M_{\odot}$, about six times larger 
than those obtained from single epoch viral black hole mass estimates. Furthermore, 
\citet{2008MNRAS.386L..15D} suggested that if the BLR has a 
disk-like geometry as opposed to a  spherical geometry, the geometrical 
factor ($f$) can fully account for the observed mass deficit in NLSy1 
galaxies and $\xi_{\mathrm{Edd}}$ turns out to be similar to 
BLSy1 galaxies \citep[see also][]{2016IJAA....6..166L}.

Arguments in favor and against the face-on view of NLSy1 galaxies
are available in the literature. Evidences in favor of the face-on view include small 
widths of the Balmer line due to projection effect 
\citep{1985ApJ...297..166O,2004MNRAS.352..823B}, anisotropic emission of Fe {\small II} \citep{2001ApJ...558..553M}, and detection of $\gamma$-rays by 
{\it Fermi} Gamma-ray space telescope in about half a dozen of the radio loud 
NLSy1 galaxies \citep{2010ApJS..188..405A}. Evidence against the face-on view come 
from the study of polarization observations by \citet{2005MNRAS.359..846S}. 
Furthermore, studying a sample of radio-loud NLSy1 galaxies, \citet{2006AJ....132..531K} 
and \citet{2008ApJ...685..801Y} could not confirm the face-on view of 
NLSy1 galaxies. An estimate of the inclination angle ($i$) of NLSy1 galaxies can be obtained using 
\begin{align}
f & = & \frac{GM_{\mathrm{BH, \sigma_{*}}}}{R_{\mathrm{BLR}} \mathrm{FWHM^2_{H\beta}}} 
	& = & (\sin^2 i + \sin^2 \omega)^{-1}
\end{align}
where $\omega$ is the half-opening angle of 
BLR geometry, $\omega=90\degree$ for a spherical geometry. 

Considering a flat BLR ($\omega=0$), we estimated $i$ for all of the objects with $f>1$. Figure \ref{Fig:inc} shows the distribution of inclination for different $M_{\mathrm{BH, \sigma_{*}}}$. The mean of the distributions are $26\degree$, $46\degree$, and $50\degree$ for $M_{\mathrm{BH, \sigma_{*}}}>10^{7.5} \, M_{\odot}$,  $M_{\mathrm{BH, \sigma_{*}}}=10^{6.8-7.5} \, M_{\odot}$ and $M_{\mathrm{BH, \sigma_{*}}}<10^{6.8} \, M_{\odot}$, respectively, though all distributions have large spreads in inclination. For the purpose of comparison, in the same way, we also calculated $i$ for the BLSy1 galaxies with $M_{\mathrm{BH, \sigma_{*}}}>10^{7.5} \, M_{\odot}$ and similar luminosity like the NLSy1 galaxies. This distribution is plotted with shaded color. The average value in the case of BLSy1 galaxies is 41$\degree$ which is larger than the average inclination of NLSy1 galaxies with $M_{\mathrm{BH, \sigma_{*}}}>10^{7.5} \, M_{\odot}$. This result is consistent with \citet{2002ChJAA...2..487Z} who estimated inclination angles of about 20 BLSy1 and 50 NLSy1 galaxies and found that the latter have systematically lower inclination angles than the former. Thus, the idea that the narrow width of the emission line in NLSy1 is due to the orientation effect seems to be true.

\begin{figure}
\centering
\resizebox{8cm}{6.0cm}{\includegraphics{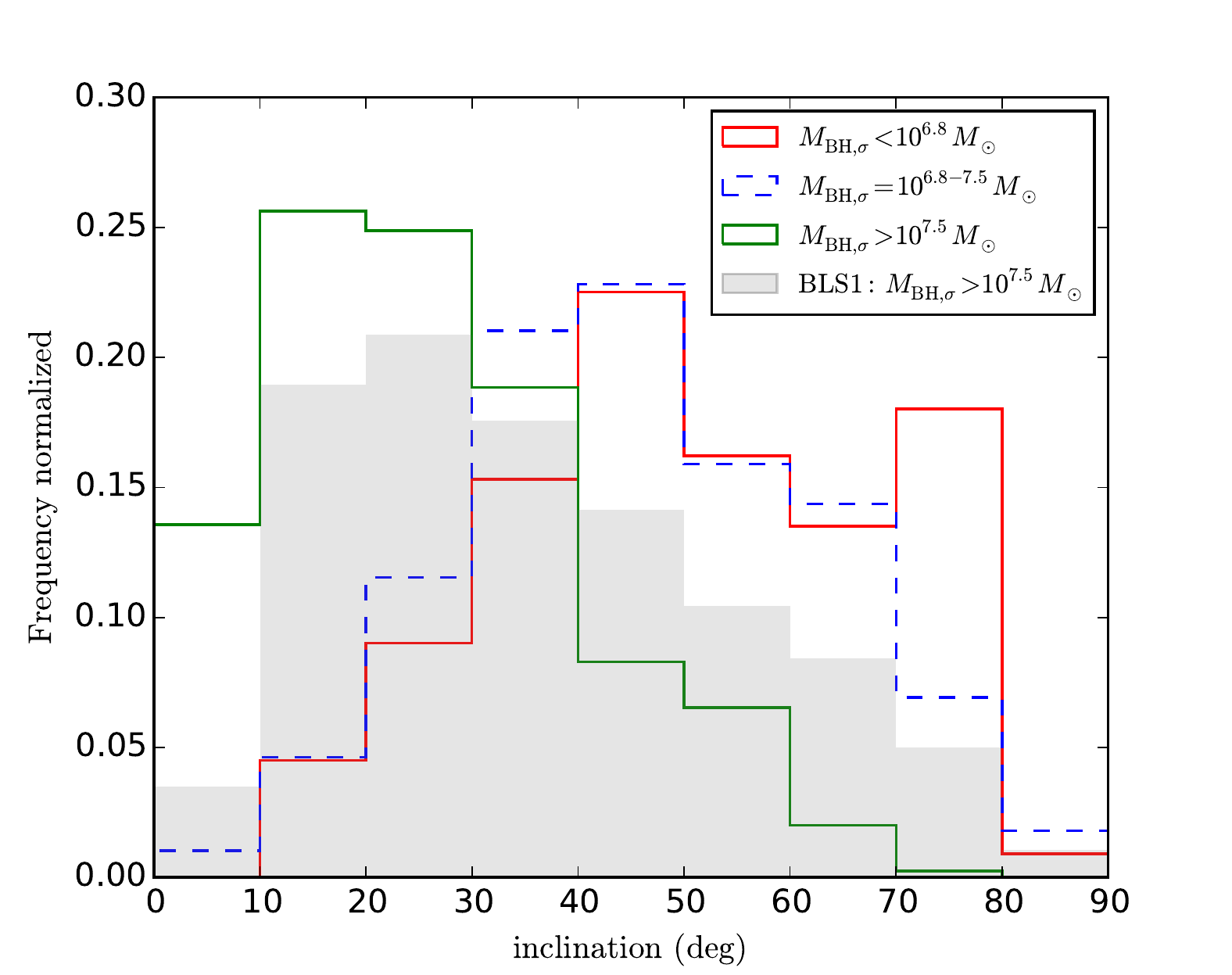}}
\caption{Distribution of inclination in degree for a range of $M_{\mathrm{BH, \sigma_{*}}}$. The shaded area represents the inclination distribution of BLSy1 having $M_{\mathrm{BH, \sigma_{*}}}>10^{7.5} \, M_{\odot}$ and similar luminosity like NLSy1. }\label{Fig:inc}. 
\end{figure}

%======================================================
%======================  Section ======================
%======================================================

\section{Summary}\label{sec:summary}

Large-scale surveys are the most effective ways to have a complete census of 
the different classes of AGNs. We have carried out a 
systematic search for new NLSy1 galaxies in SDSS DR12 by using a sample of 68,859 objects having $z<0.8$ and a median of S/N $>2 \, \mathrm{pixel^{-1}}$. For
this, we have developed procedures to carefully subtract the host galaxy 
contribution, taking into account the Fe {\small II} emission and
extracting the nuclear continuum contribution, and then fitting the 
emission lines to extract crucial emission line parameters. Applying
our developed procedure to SDSS DR12 database, we compiled a
new sample of NLSy1 galaxies. The major findings of this work are as follows.

\renewcommand\labelitemii{$\square$}
\begin{enumerate}
\item Adopting the criteria of  FWHM(H$\beta$) $\le 2200 \mathrm{\,km \, s^{-1}}$
and [O {\small III}]/H$\beta<3$ to our derived emission line parameters of ``QSO'' sources in SDSS DR12, we arrived at a new sample of 11,101 NLSy1 
galaxies. This is about five times larger than the 2011 NLSy1 galaxies known 
previously from SDSS DR3 (ZH06). % \citep{2006ApJS..166..128Z}.

\item The broad Balmer component of the majority of the NLSy1 galaxies can be best-fitted with a Lorentzian than a Gaussian profile reaffirming the claim made in the literature with a smaller sample.

\item A total of 68,859 objects with $z$ $<$ 0.8 and a median of S/N $>2 \,\mathrm{pixel^{-1}}$ has been studied. Of this, 11,101 are NLSy1
galaxies, which indicate that $\sim$ 16\% of AGNs are NLSy1 galaxies. The absolute magnitude of this sample peaks at around $M_{g} \simeq -22$ mag. 
 
\item The monochromatic luminosity at 5100 $\mathrm{\AA}$ is found to be strongly 
correlated with luminosity of H$\beta$, H$\alpha$, and [O {\small III}] lines, which is in agreement with the results already available in the literature \citep[e.g.,][]{2005ApJ...630..122G}.    

\item The strength of Fe {\small II} emission in NLSy1 galaxies is a factor of two larger
than BLSy1 galaxies. The $R_{4570}$ distribution peaks at 0.64 in NLSy1 galaxies and 
0.38 in BLSy1 galaxies. About 0.5\% of NLSy1 galaxies shows strong Fe {\small II} emission with $R_{4570}>2$.

\item The equivalent width of Fe {\small II} and H$\beta$ is found to be strongly 
correlated with the luminosity of H$\beta$ and the monochromatic luminosity at 
5100 $\mathrm{\AA}$, which agrees with ZH06 and partially with \citet{2001A&A...372..730V}. An anti-correlation between $L_{\mathrm{H\beta}}$ with $R_{5007}$ is found, however, no correlation with $R_{4570}$ is present.

\item In our new sample of NLSy1 galaxies, 555 (5\%) sources are detected in the FIRST survey. Such a low fraction of radio emitting NLSy1 galaxies (7\%) is also reported by ZH06. Thus, NLSy1 galaxies do show the radio-loud/radio-quiet dichotomy seen in the 
quasar category of AGNs. The distribution of radio power distribution peaks at 40.5 erg s$^{-1}$ and shows a strong correlation with [O {\small III}] luminosity at $z<0.2$.

\item About 17\% of our sample of NLSy1 galaxies are detected in the
soft X-ray band by \textit{ROSAT}. The soft X-ray flux distribution peaks at low flux, 
around $10^{-12} \mathrm{erg\, s^{-1}\, cm^{-2}}$, while the number falls 
rapidly for high flux.

\item The strong difference in  $M_{\mathrm{BH}}$ (calculated using viral methods)
and Eddington ratio are found between NLSy1 and BLSy1 galaxies. NLSy1 galaxies have low 
 $M_{\mathrm{BH}}$ and high Eddington ratio compared to BLSy1 galaxies. 

\item Electron density in the NLR of NLSy1 galaxies is found to be low and widely 
scattered compared to BLSy1 galaxies. About 25\% of NLSy1 and 15\% of BLSy1 galaxies have low density $n_e<140\,\mathrm{cm}^{-3}$. This is against the finding of \citet{2007ApJ...670...60X} who found the BLSy1 galaxies not to have $n_e<140\,\mathrm{cm}^{-3}$. Thus, the ``zone of avoidance'' in density found for BLSy1 galaxies by \citet{2007ApJ...670...60X} does not exist.

\item Stellar velocity dispersion ($\sigma_{*}$) has been obtained for 1789 
NLSy1 galaxies. A positive correlation has been found between $\sigma_{*}$ and 
 $M_{\mathrm{BH}}$ and a slightly weaker correlation is present 
between $M_{\mathrm{BH}}$ and FWHM of the [O {\small III}] narrow component. 

\item The average inclination of NLSy1 galaxies is lower compared to BLSy1 galaxies. This suggests an inclination as the main geometrical parameter responsible for the black hole mass deficit in NLSy1 galaxies.

\end{enumerate}

\acknowledgments
We are grateful for the comments and suggestions by the anonymous referee, which helped to improve the manuscript.
This work has made use of SDSS spectroscopic data. Funding for SDSS-III has been provided by the Alfred P. Sloan Foundation, the Participating Institutions, the National Science Foundation, and the U.S. Department of Energy Office of Science. The SDSS-III website is http://www.sdss3.org/. SDSS-III is managed by the Astrophysical Research Consortium for the Participating Institutions of the SDSS-III Collaboration including the University of Arizona, the Brazilian Participation Group, Brookhaven National Laboratory, Carnegie Mellon University, University of Florida, the French Participation Group, the German Participation Group, Harvard University, the Instituto de Astrofisica de Canarias, the Michigan State/Notre Dame/JINA Participation Group, Johns Hopkins University, Lawrence Berkeley National Laboratory, Max Planck Institute for Astrophysics, Max Planck Institute for Extraterrestrial Physics, New Mexico State University, New York University, Ohio State University, Pennsylvania State University, University of Portsmouth, Princeton University, the Spanish Participation Group, University of Tokyo, University of Utah, Vanderbilt University, University of Virginia, University of Washington, and Yale University.

S.R. thanks Neha Sharma (ARIES, India) for carefully reading the manuscript.  

 \bibliographystyle{apj}
 \bibliography{ref}

\end{document}